\pdfoutput=1
\UseRawInputEncoding
\documentclass[aps,pra,reprint,superscriptaddress,showpacs]{revtex4-1}

\usepackage{graphicx}
\usepackage{dcolumn}
\usepackage{bm}
\usepackage[fleqn]{amsmath}
\usepackage{amsfonts,amssymb} 
\usepackage{hyperref}
\usepackage{url}

\makeatletter
\renewcommand{\maketag@@@}[1]{\hbox{\m@th\normalsize\normalfont#1}}%
\makeatother

\begin{document}
\preprint{APS/123-QED}

\title{Topological Robust Corner States of a Two-Dimensional Square Lattice with \texorpdfstring{$\mathbf C_{\mathbf 4}$}{} Symmetry in Fully Coupled Dipolar Arrays}
\author{Chen Luo}
\email{nc\_lc@email.ncu.edu.cn}
\affiliation{Department of Physics, Nanchang University, Nanchang 330031, China.}  

\author{Xiang Zhou}%
\affiliation{Department of Physics, Nanchang University, Nanchang 330031, China.}  
\author{Hui-Chang Li}
\affiliation{Department of Physics, Nanchang University, Nanchang 330031, China.}  

\author{Tai-Lin Zhang}
\affiliation{Department of Physics, Nanchang University, Nanchang 330031, China.}  

\author{Yun Shen}
\email{Corresponding author:shenyun@ncu.edu.cn}
\affiliation{Department of Physics, Nanchang University, Nanchang 330031, China.}  

\author{Xiao-Hua Deng}
\email{dengxhua@gmail.com}
\affiliation{Institute of Space Science and Technology, Nanchang University, Nanchang 330031, China.}  

\begin{abstract}

Higher-order topological insulators(HOTIs) is an exciting topic. We constructed a square lattice dipole arrays, it supports out-of-plane and in-plane modes by going beyond conventional scalar coupling. In-plane modes naturally break $\mathrm C_{4}$ symmetry, we only studied the out-of-plane modes that maintain $\mathrm C_{4}$ symmetry. Due to the slowly decaying long-range coupling, we consider its fully coupled interactions by using the lattice sums technique and combined with the coupled dipole method (CDM) to study its topological properties in detail. Interestingly, even when the full coupling is considered, the topological properties of the system remain similar to those of the 2D Su-Schrieffer-Heeger(SSH) model, but very differently, it supports robust zero-energy corner states (ZECSs) with $\mathrm C_{4}$ symmetry, we calculate the bulk polarization and discuss in detail the topological origin of the ZECSs. The lattice sums technique in the article can be applied to arbitrary fully coupled 2D dipole arrays. The materials we used can be able to confine light into the deep subwavelength scale, it has a great potential in enhancing light-matter interactions in the terahertz (THz) range.

\end{abstract}
\maketitle


\section{\label{sec1}INTRODUCTION}

The discovery of topological insulators (TIs) is one of the most exciting advances in condensed matter physics. It has counterintuitive properties of body insulation and edge transport. The discovery of HOTIs\cite{A16} further enriched the connotation of topological phases of matter, its biggest feature is an n-D topological insulator supporting at least (n-2)-D topological protection states. Unlike the Chern insulators and valley insulators protected by nonzero Berry curvature, this behavior needs to be explained by a topological invariant based on the wannier center. HOTIs have been ubiquitously extended to all areas of classical wave physics, from microwaves \cite{A22}, acoustic waves \cite{A21}, circuits\cite{A19}, and even mechanical\cite{A20}, heat transfer\cite{A18} and so on, this proves the existence of HOTIs both theoretically and experimentally. Recently, the effect of long-range interactions (LRIs) on higher-order topological phases (HOTPs) has attracted extensive interest \cite{A15,A26,A27,B1}. In contrast to the previous systems where only nearest-neighbor (NN) coupling was considered, it was found that LRIs can support new HOTPs, such as type-II and type-III corner states that do not exist in NN coupling systems \cite{A15}. Under the LRIs, type II and III corner states are separated from the edge states \cite{A15}, so they are not always strictly localized to the corners of the system. Type-I corner states are widely found in NN coupling systems and long-range coupling systems and are also called zero-energy corner states(ZECSs) because of the generalized chiral symmetry leading to it being localized at zero-energy \cite{A15}.

In the standard 2D SSH model, due to the protection of $\mathrm C_{4}$ symmetry, the degeneracy of the 2nd and 3rd bands is hardly broken and thus no zero-energy band gap appears\cite{A29}, so it doesn't support the type-I corner states. In order to break the degeneracy so that the type-I corner states appears in the zero-energy band gap, it was found that a square lattice without time-reversal symmetry (TRS) could be constructed to achieve this, but this requires an additional magnetic field\cite{A30}, or by applying geometric modulation methods to break the $\mathrm C_{4}$ symmetry, such as twisting the unit-cell\cite{A31} and removing some scatterers\cite{A32}. Surprisingly, in some previous studies\cite{A36,A28}, the in-gap corner states also appeared in some analogs of the standard 2D SSH model for optical platforms that maintain $\mathrm C_{4}$ symmetry and TRS. The reason is that the practical optical system differs from the ideal tight-binding model, in which there are always LRIs that are not zero, and some recent works have theoretically elaborated the basic reasons for the appearance of this phenomenon\cite{A26,A27}.

In this regard, we consider a practical analog based on the promoted 2D SSH model, which also maintains the $\mathrm C_{4}$ symmetry and the TRS. Specifically, we constructed a square lattice dipole array composed of silicon carbide (SiC), relative to the previous research\cite{A26},for rigorous considerations and a natural mathematical generalization of similar works \cite{A50,A47}, we consider the interaction of the lattice point with all lattice points except itself. Meanwhile, due to the natural $\mathrm C_{4}$ symmetry breaking of the in-plane modes\cite{A41},we only studied the out-of-plane modes that maintain $\mathrm C_{4}$ symmetry and used a lattice sums technique called Ewald method\cite{A10,A13,A14} to calculate the full coupling of the system. Under the LRIs, the 2nd and 3rd bands of the system are shifted upward, which makes the robust type-I corner states appear in the zero-energy band gap, and we calculate the bulk polarization in the fully coupled case, which still accurately captures the topological phase transition of the system. Our work provides new insights into the implementation of HOTPs in simple lattices, and the lattice sums technique of which can be widely applied to other 2D periodic dipole arrays, paving the way for highly integrated optical devices utilizing exotic topological photonic phases.

The paper is organized as follows. In Sec.\ref{sec2}, the CDM is introduced and combined with the Ewald method to construct the characteristic equations and calculate the bulk bands. In Sec.\ref{sec3} and Sec.\ref{sec4}, the topological states of the system and the corresponding bulk moments are calculated, and the origin of the robust corner states is discussed in detail. Finally, a brief discussion and conclusion is given in Sec.\ref{sec5}.

\section{\label{sec2}COUPLED DIPOLE MODEL }

\subsection{\label{subsec:A}Lattice sum of Green’s function }
The structure discussed in this paper is shown in Fig. 1 (a), and the gray dashed box is the dividing line between the unit cells (UCs), each of which contains four spherical SiC nanoparticles with radius $ a$. The lattice constant of the UCs is $ d$, and the distance between the two particles in the x- and y-directions of the UCs is $ 2c$. The red numbers are the numbers of the particles in the UCs, and the characteristic equations will be constructed according to this order later.Unlike the spherical metal nanoparticle arrays that form dispersion relations based on the interaction of localized surface plasmon polaritons\cite{A41,A42,A44,A45,A46,A48,A49,A50,A51}, the SiC nanoparticle arrays are formed by phonon polaritons to form coupling effects between each other\cite{A47,A53}, but they can both be based on the CDM to describe the dipole-dipole interactions between lattice points \cite{A39}.Let's start with the coupled dipole equation:
\begin{align}
\frac{1}{\alpha(\omega)} \mathbf{p}_i=\frac{k^2}{\epsilon_0} \sum_{j \neq i} \mathrm{\mathbf G}\left(\mathbf{r}_{i j}, \omega\right) \mathbf{p}_j,
\label{eq1}
\end{align}
where $ \mathbf{p}_i=(p_i^x, p_i^y, p_i^z)^{\mathrm{T}}$ is a $ 3\times1$ vector representing the dipole moment at position $ i$, and $ x, y, z$ represents the three fundamental directions.$ \alpha(\omega)$ is a scalar, representing the polarizability, and also has the relation $ \mathbf{p}_i=\alpha\mathbf{E}_i$, $ \mathbf{E}_i$ represents the electric field vector at position $ i$.$ \epsilon_0$ is the vacuum permittivity, we set the relative permittivity of the background to 1, and $ k$ represents the wave vector in the vacuum.The summation represents the position $ j$ traversing all positions except $ j=i$, so the meaning of Eq.(\ref{eq1}) is that the electric field at position $ i$ is related to the electric field at all positions except itself, and the coupled dipole system can be used as a platform for implementing the topology.

The $ \mathbf{G}$ in Eq.(\ref{eq1}) is the dyadic Green's function(DGF), and the specific expression is:
\begin{scriptsize}
\begin{align}
\label{eq2}
\notag
&\mathbf{G}\left(\mathbf{r}_{i j}, \omega\right)\\ &=\left\{\hat{\mathbf{I}}+\frac{1}{k^2}\nabla\otimes \nabla \right\} \frac{e^{-i kR}}{4 \pi R} \\
&=\frac{e^{-i k R}}{4 \pi R}\left\{\hat{\mathbf{I}}-\frac{i k R+1}{k^2 R^2} \cdot \hat{\mathbf{I}}+\frac{3+3 i k R-k^2 R^2}{k^2 R^2} \cdot \frac{\mathbf{r}_{i j} \otimes \mathbf{r}_{i j}}{R^2}\right\}\notag,
\end{align}
\end{scriptsize}
where $ \hat{\mathbf{I}}$ is the unit tensor,$ \mathbf{r}_{i j}=\mathbf{r}_i-\mathbf{r}_j$, $ \mathbf{r}_i,\mathbf{r}_j$ is the position vector at position $ i, j$ , $ R=|\mathbf{r}_{i j}|$. $ \mathbf{G}$ is a tensor of $ 3 \times3$.
Since we only consider out-of-plane modes (Z-modes), Eq.(\ref{eq2}) can be reduced to:$
\textstyle \mathbf{G}=\frac{e^{-i k R}}{4 \pi R}\left\{1-\frac{i k R+1}{k^2 R^2}\right\}$.

Beyond the quasistatic approximation \cite{A40}, the polarizability $ \alpha(\omega)$ in Eq.(\ref{eq1}) is expressed specifically as \cite{A53}:
\begin{align}
\alpha(\omega) &=\frac{\alpha_{\mathrm{QS}}(\omega)}{1-i \frac{\mathrm{k}^3}{6 \pi \epsilon_0} \alpha_{\mathrm{QS}}(\omega)},
\label{eq3}
\end{align}
where $ \alpha_{\mathrm{QS}}(\omega)$ represents the polarizability under the quasistatic approximation:
\begin{align}
\alpha_{\mathrm{QS}}(\omega) &=4 \pi a^3 \epsilon_0 \frac{\epsilon(\omega)-1}{\epsilon(\omega)+2 },
\label{eq4}
\end{align}
the $a$ in Eq.(\ref{eq4})is the radius of the spherical nanoparticle, $\epsilon(\omega)$ represents the relative permittivity of SiC, characterized by the Lorentz model $\epsilon(\omega)=\varepsilon_{\infty}\left(1+\frac{\omega_L^2-\omega_T^2}{\omega_T^2-\omega^2-i\omega \gamma}\right)$,where $\varepsilon_{\infty}$ = 6.7 is the high frequency limit of the dielectric constant,$\omega_{T}$ = 790 $\mathrm{cm}^{-1}$ is the transverse optical phonon frequency,$\omega_{L}$ = 966 $\mathrm{cm}^{-1}$ is the longitudinal optical phonon frequency and $\gamma$ = 2 $\mathrm{cm}^{-1}$ relates to the nonradiative damping rate in the material.

Following the idea of Eq.(\ref{eq1}) and combining it with Bloch theorem, we construct the characteristic equation considering the fully coupled case:
\begin{align}
\frac{1}{\alpha}\left(\begin{array}{cccc}
\mathbf{p}_0^1 \\[1.0mm]
\mathbf{p}_0^2 \\[1.0mm]
\mathbf{p}_0^3 \\[1.0mm]
\mathbf{p}_0^4
\end{array}\right)=\frac{k^2}{\epsilon_0}\mathcal{M}
\left(\begin{array}{cccc}
\mathbf{p}_0^1 \\[1.0mm]
\mathbf{p}_0^2 \\[1.0mm]
\mathbf{p}_0^3 \\[1.0mm]
\mathbf{p}_0^4
\end{array}\right),
\label{eq5}
\end{align}
the $ \mathcal{M}$ in Eq.(\ref{eq5}) is the eigenmatrix:
\begin{widetext}
\begin{equation}
\mathcal{M}=
\left(\begin{array}{cccc}
\Sigma^{\prime} \mathbf{G}(\vec{R}) \cdot t & \Sigma \mathbf{G}\left[\vec{R}+\vec{r}_{21}\right] \cdot t & \Sigma \mathbf{G}\left[\vec{R}+\vec{r}_{31}\right] \cdot t & \Sigma \mathbf{G}\left[\vec{R}+\vec{r}_{41}\right] \cdot t \\[2mm]
\Sigma \mathbf{G}\left[\vec{R}+\vec{r}_{12}\right] \cdot t & \Sigma^{\prime} \mathbf{G}(\vec{R}) \cdot t & \Sigma \mathbf{G}\left[\vec{R}+\vec{r}_{32}\right] \cdot t & \Sigma \mathbf{G}\left[\vec{R}+\vec{r}_{42}\right] \cdot t \\[2mm]
\Sigma \mathbf{G}\left[\vec{R}+\vec{r}_{13}\right] \cdot t & \Sigma \mathbf{G}\left[\vec{R}+\vec{r}_{23}\right] \cdot t & \Sigma^{\prime} \mathbf{G}(\vec{R}) \cdot t & \Sigma \mathbf{G}\left[\vec{R}+\vec{r}_{43}\right] \cdot t \\[2mm]
\Sigma \mathbf G\left[\vec{R}+\vec{r}_{14}\right] \cdot t & \Sigma \mathbf{G}\left[\vec{R}+\vec{r}_{24}\right] \cdot t & \Sigma \mathbf{G}\left[\vec{R}+\vec{r}_{34}\right] \cdot t &  \Sigma^{\prime} \mathbf{G}(\vec{R}) \cdot t
\end{array}\right),\label{eq6}
\end{equation}
\end{widetext}
where $  t =e^{i \vec{q} \cdot \vec{R}},\Sigma =\sum_{(m, n) \in Z},\Sigma^{\prime} =\sum_{(m,n) \neq (0,0)}$;$ \vec{q}=q_x{\hat{x}} + q_y {\hat{y}}$ is the Bloch wave vector; $ \vec{R}=m d\cdot \hat{x}+n d \cdot \hat{y}$ denotes the vector between any two unit cells,$ m,n$ is the summation parameter ($ m,n\in{Z}$);$ (\mathbf{p}_0^1,\mathbf{p}_0^ 2 ,\mathbf{p}_0^3 ,\mathbf{p}_0^4)^{\mathbf{T}}$ denotes the dipole moment at $ (m,n)=(0,0)$, and the superscript serial number denotes the particle number within the unit cell,as shown in Fig.\ref{fig1}(a).$ \vec{r}_{u v}=\vec{r}_u-\vec{r}_v,(u, v=1, 2, 3, 4$ represent the particle numbers inside the unit cell), under the particle numbers shown in Fig.\ref{fig1}(a) there are:$ \vec{r}_1=(c, c),\vec{r}_2=(-c,-c),\vec{r}_3=(-c, c),\vec{r}_4=(c,-c)$, the coordinate center is the center of the unit cell, while $ c=(\beta d)/{4}$,$ \beta \in(0,2) $, the lattice becomes a uniform lattice when $ \beta=1$.

In the case of considering only Z-modes, the eigenmatrix $ \mathcal{M}$ reduces to a matrix of $4\times 4$, which is similar to the bulk Hamiltonian in the tight binding method.It can be noticed that each term inside $ \mathcal{M}$ is an infinite series. To avoid brute-force summation, we need to find an effective lattice sums technique to accelerate the convergence of these infinite series.The Ewald method \cite{A10,A13,A14} has been successfully applied to the analysis of electromagnetic calculations of various active periodic structures, which converts the slowly converging original Green's function series into a summation of real space terms and k-space terms with rapid (Gaussian) convergence rates,and only a few terms are needed to achieve excellent convergence,so here we introduce it to the calculation of coupling coefficients in passive topological systems.Observing the matrix $ \mathcal{M}$, we can find that most of the terms inside are repeated, we only need to compute $ \Sigma \mathbf{G}[\begin{array}{l}\vec{R}+\vec{r}_{u v}\end{array}] \cdot t,\Sigma^{\prime} \mathbf{G}(\vec{R}) \cdot t $, in the Appendix \ref{smsec1}),we derive the Ewald series form of these two terms, and verify their accuracy in detail.

\subsection{\label{subsec:B}Bulk bands}
We observe that Eq.(\ref{eq5}) is a nonlinear equation since both sides of the equal sign contain variables $ \omega$ and therefore it is not possible to directly find the eigenvalues and eigenvectors of the matrix $ \mathcal{M}$.To solve this problem in previous studies, some use the so-called complex root searching method \cite{A50,A40}, which has the core idea of finding a suitable $ \omega$ at each point of the Brillouin zone path such that $ \mathrm{det}[\mathcal{M}(q_x,q_y,\omega)-\mathbf{\hat I}/{\alpha(\omega)}]=0$ holds, but when the eigenmatrix $ \mathcal{M}$ is large, especially when considering an open-boundary structure, this method is abandoned due to extremely large computer memory costs and a long root finding time;some are using another method called eigen-response theory \cite{A37,A38,A43,A45,A46,A53},its core idea is to calculate the eigenvalue $\lambda$ of the matrix $ \hat {\mathbf I}/{\alpha(\omega)}-\mathcal{M}(q_x,q_y,\omega)$ in the $ (q_x,q_y,\omega)$ parameter space, and to plot the heat map of $\mathrm{Im}(1/\lambda)_{\mathrm{max}}$,this heat map can reveal the dispersion relation of the system.It is commonly used because it easier to implement and has a smaller computational overhead. However, since this method is not convenient for calculating topological invariants, it has not been used in this paper to calculate the bulk bands and bulk polarization.Fortunately, we found a simple and effective method which is implemented by setting all $ \omega$ on the right hand side(RHS) of Eq.(\ref{eq5})to a fixed value, but it should be noted that all $ \omega$ on the left hand side (LHS) of Eq.(\ref{eq5}) should be treated as variables, and these steps are called linearization \cite{A40,A47}, so that we can directly solve for the eigenvalues and eigenvectors of the $\mathcal{M}$.Specifically, we set the radius of the SiC nanoparticles to 100 $\mathrm{nm}$, and keep it constant throughout the rest of the paper. Since the radius of the nanoparticles does not affect the topological properties of the system \cite{A40}, this setting can be considered reasonable. Under this radius parameter, the Dirac frequency $ \omega_\mathrm{D}$ is $928.5 \mathrm{cm^{-1}}$\cite{A53}, so we fix all the $ \omega$ on the RHS of Eq.(\ref{eq5}) as $\omega_{\mathrm{D }}$.The $\omega_\mathrm{D }$ is a real number, this avoids a divergence problem in the lattice sums \cite{A52}.
\begin{figure}[!htbp]
\centering
\includegraphics[width=0.95\linewidth]{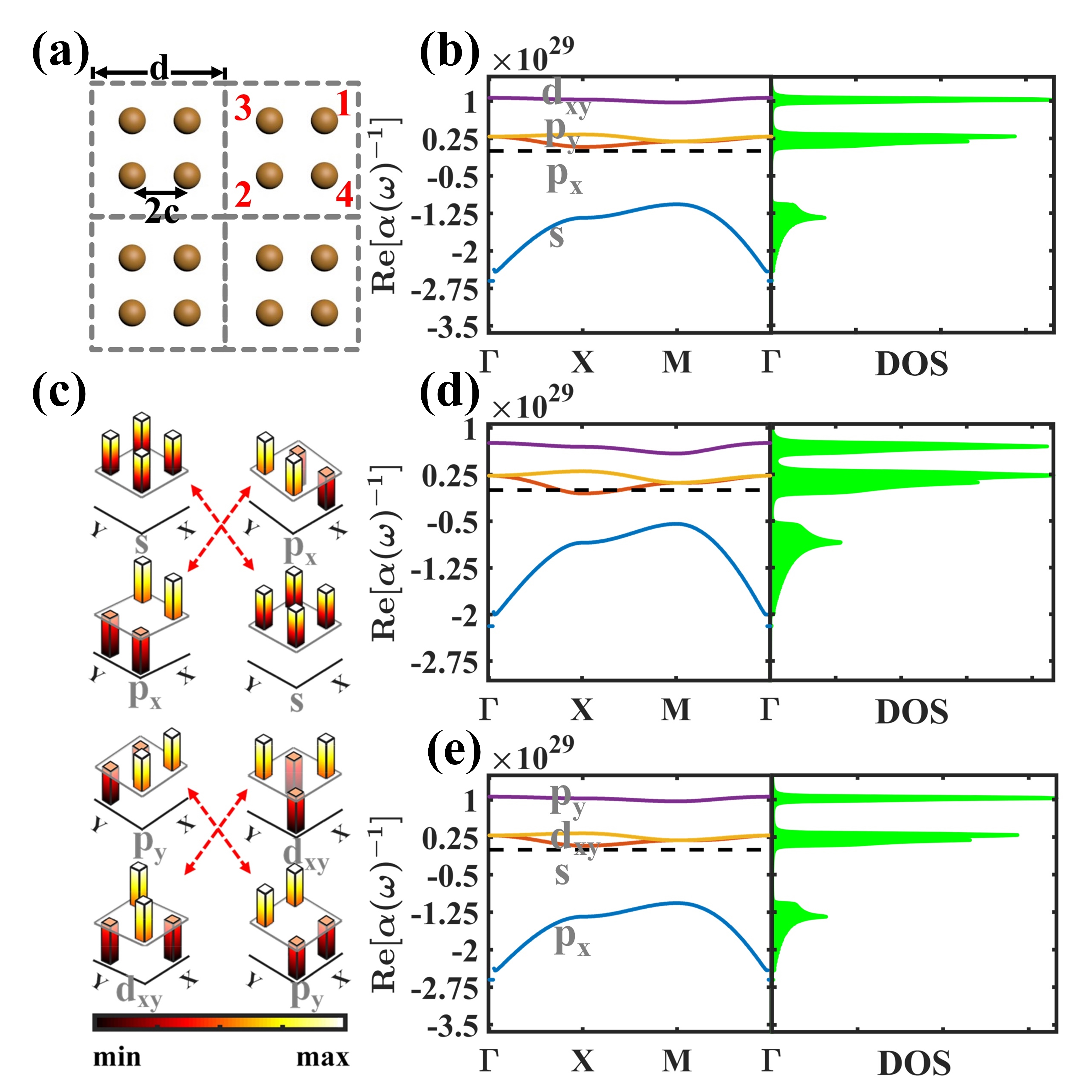}
\caption{(a) Schematic diagram of the structure, where the gray dotted line is the partition line of the unit cell, d is the lattice constant, 2c is the distance between the centers of two adjacent particles in the x or y direction of the intra-cell, and the red numbers are the numbers of the particles of the intra-cell, which are used for the derivation of the Eq.(\ref{eq5}).The left panels of (b), (d), and (e) are the bulk bands considering the fully coupled case,the lattice constant d=15a, $\beta$ is 0.7, 1.2, and 1.3, respectively, and the black dotted lines are for the case where the y-axis is 0, which is related to the appearance of the corner states,the  bands do not intersect the dotted lines in both the 0.7 and 1.3 cases, while  intersect the dotted lines in the 1.2 case; the right panel is the DOS under the corresponding parameters.The left panel of (c) shows the four field distributions of the energy band near the $ \mathbf{X}$ point at $\beta$=0.7, and the right panel shows the four field distributions of the energy band near the $ \mathbf{X}$ point at $\beta$=1.3. From 0.7 to 1.3, the energy bands are inverted, indicating that there is a topological phase transition in it.}
\label{fig1}
\end{figure}

Under the assumption of linearization we calculated the band structure of the periodic structure along the path of high symmetry points in the first Brillouin zone as shown in Fig.\ref{fig1}(b),(d),and (e).With the idea of Ref.\cite{A40}, for simplicity, we only computed $ 1/\alpha(\omega)$ and took it as an eigenvalue. In fact, this consideration is reasonable because the topological properties presented by such a band structure are exactly the same as those of the band structure drawn with $\omega$ as an eigenvalue, and the band structure in both cases only differs geometrically by flipping up and down.Also, we only considered the real part of the eigenvalues, since the imaginary part is often three orders of magnitude smaller than the real part, all of which we will demonstrate in Appendix \ref{smsec5}.The parameters for the band structure plotting are set as follows: the lattice constant $\mathrm{d=15a}$ (which remains unchanged throughout the rest of the paper), and the geometric parameters $ \beta$ are 0.7, 1.2, and 1.3 in Fig.\ref{fig1}(b), (d), and(e), respectively.The right panels of (b), (d), and (e) are the density of states (DOS)\cite{A23} calculated using the right vector and its corresponding eigenvalues, and the small quantities used in the calculations are three orders of magnitude smaller than the eigenvalues in order to ensure good convergence, and all the DOS capture the features of the corresponding energy bands.As in the standard 2D SSH model, when $ \beta=1$, the $ \mathbf{ M}$ point is quadruply degenerate, i.e., it is a phase transition point, as is confirmed later in this paper.The black dotted line in the energy band diagram is the case where the y-axis value is 0 (corresponding to $\omega/\omega_\mathbf{D}=1$), which is closely related to the appearance of the corner states.In Fig.\ref{fig1}(b) and (e) the energy bands do not intersect the black dotted line, and we also calculate the field distribution of the four energy bands near the $\mathrm{\mathbf X}$ point in these two cases, as shown in Fig.\ref{fig1}(c) the left panel is the case of $ \beta=0.7$ the right panel is the case of $\beta=1.3$. It can be found that the energy bands are inverted from 0.7 to 1.3, which indicates that there is a phase transition, and the results of the calculation of the bulk polarization later again support this point.The energy band intersects the black dotted line in Fig.\ref{fig1}(d) and the band does not intersect it in (e), implying that the ZECSs first appear between 1.2 and 1.3, and we determine this specific value in the discussion in Sec.\ref{sec4}. Going beyond the linearization assumption, in Appendix \ref{smsec4} we investigate the topological properties of the open boundary structure using eigen-response theory, which is consistent with that described earlier in the paper.

\section{\label{sec3}EDGE STATES}
To further investigate the topological properties of the system, we calculated the projection bands of the fully coupled nanoribbons with open boundaries in the x-direction and periodic boundaries in the y-direction.By considering 20 UCs in the x-direction and one in the y-direction, following the idea of constructing the eigenmatrix in Sec.\ref{sec2}, an eigenmatrix of size $80 \times 80$ can be constructed when only the Z-modes are considered, and after linearization, its eigenvalues and eigenvectors can be directly derived.Note that unlike Sec.\ref{sec2}, we are not using the lattice sums technique here to accelerate the convergence of the infinite series because we find that the Ewald method may not be applicable to the calculation of the energy band of the open boundary structure, and therefore only the dipole-dipole interaction of the 1000th farthest supercell up and down the y-direction of the nanoribbon is considered here.In Appendix \ref{smsec1} and Appendix \ref{smsec3}, we still derive the lattice sums technique for this mixed boundary condition and plot the projection bands in the range where it holds, which present topological properties consistent with those obtained using truncated summation.In practice, it has been found that truncated summation tends to be faster than using the lattice sums technique under the same conditions, which we believe is due to the fact that the lattice sums technique involves complex special functions and more summation steps than truncated summation.
\begin{figure*}[!htbp]
\centering
\includegraphics[width=0.85\linewidth]{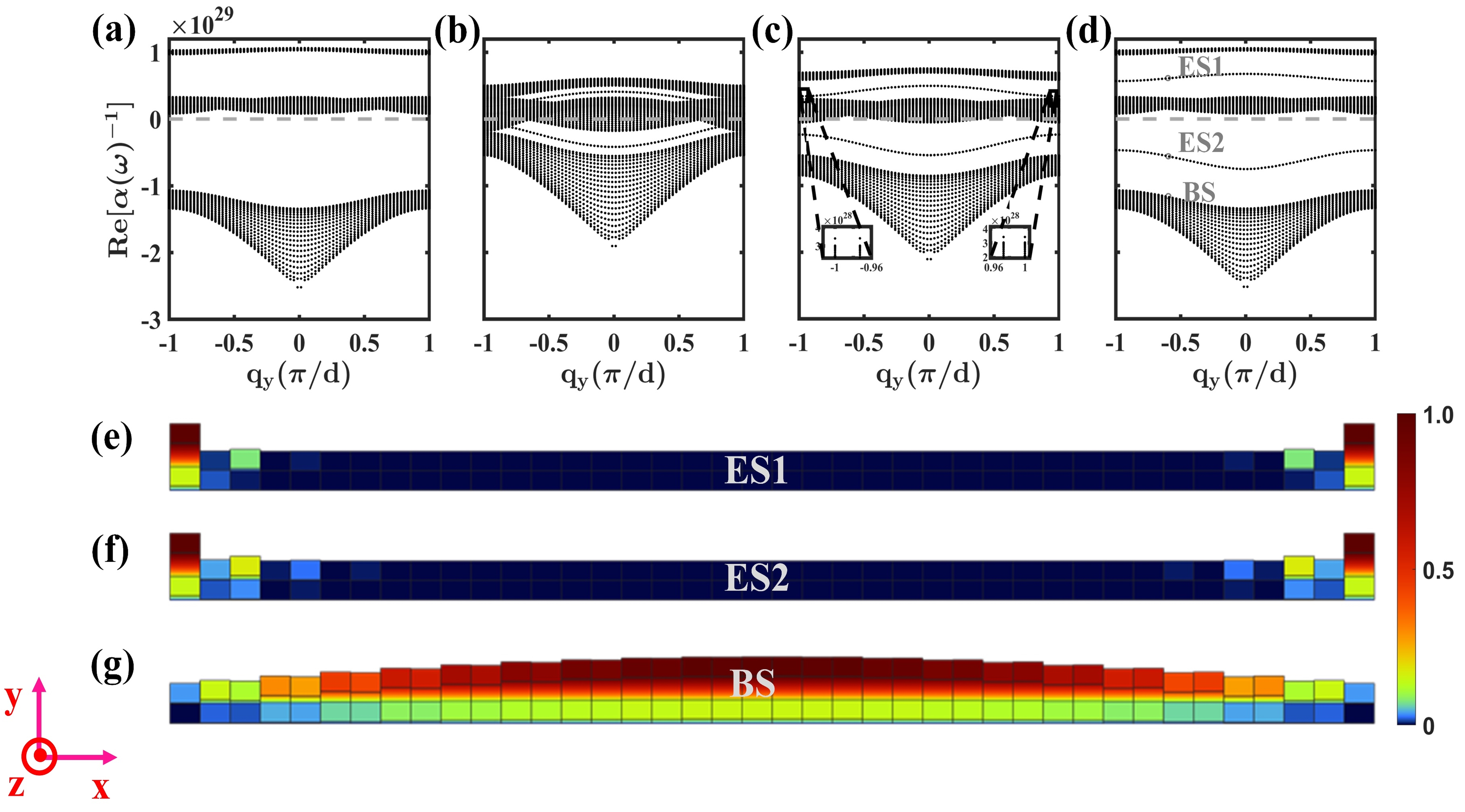}
\caption{Projection bands and dipole moment distributions of fully coupled nanoribbons plotted using truncated summation.(a)-(d) are the projection bands, and the gray dotted line is the case where the y-axis is zero, which is related to the appearance of the corner states;the lattice constants $d=15a$,$\beta$ are 0.7,1.1,1.2,1.3, respectively.According to the trend, $\beta$ is the topological phase transition point when $\beta$ is 1, the complete bandgap inner edge states appears for the first time at $\beta$ = 1.2, and the bulk does not intersect the dotted line for the first time between 1.2 and 1.3.(e-g) is the dipole moment distribution near $q_y=-0.5\pi/d$ of the (d) , ES1 and ES2 are the edge states and BS is the bulk state.}
\label{fig2}
\end{figure*}

Similar to Sec.\ref{sec2}, most of the elements in the eigenmatrix are repeated, and we only need to consider $ \sum_{n \in Z} \mathbf{G}(n d \cdot \hat{y}+\vec{r}) \cdot e^{i q_y \cdot n d}$ and $ \sum_{n \neq 0} \mathbf{G}(n d \cdot \hat{y})\cdot e^{i q_y \cdot n d}$,where $n$ is the summation parameter, which is an integer, $\vec{r}=r_x\hat{x}+r_y\hat{y}+r_z\hat{z}$ denotes the difference in position vectors between any two nanoparticles at different positions in the same nanoribbon (here $r_z=0$), and the other parameters have the same meaning as described earlier in the paper.Using the truncated summation approach while considering only the $(3,3)$ element of the DGF tensor Eq.(\ref{eq2}) in the Z-modes, these two terms can be reduced separately as:
\begin{small}
\begin{align}
\sum_{n \in {Z}} \frac{e^{-i k \mid n d \cdot \hat{y}+\vec{r} \mid}}{4 \pi|n d \cdot \hat{y}+\vec{r}|}\left\{1-\frac{i k|n d\cdot\hat{y}+\vec{r}|+1}{k^2|n d \cdot \hat{y}+\vec{r}|^2}\right\}  e^{i q_y \cdot n d},
\label{eq7}
\end{align}
\end{small}
and
\begin{small}
\begin{align}
\sum_{n \neq 0} \frac{e^{-i k|n d|}}{4 \pi|n d|}\left\{1-\frac{i k|n d|+1}{k^2|n d|^2}\right\}  e^{i q_y \cdot n d}.
\label{eq8}
\end{align}
\end{small}

We assume $ |n|\leq 1000$, and directly sum up Eqs.(\ref{eq7}) and (\ref{eq8}) to obtain the eigenmatrix of the nanoribbons and draw the energy bands and field distributions as shown in Fig.\ref{fig2}.Fig.\ref{fig2}(a-d) is the projection bands, the lattice constant is set as before, $\beta$ is set to 0.7, 1.1, 1.2, 1.3 respectively, and the gray dotted line is the case where the y-axis is zero;Fig.\ref{fig2}(e-g) are the field distributions of the nanoribbons.Observing the results we find that Fig.\ref{fig2} is able to capture more topological properties of the system than Fig.\ref{fig1}.When $\beta=0.7$, the energy band does not intersect the gray dotted line, but there is no edge states(ESs) within the band gap and the system is topologically trivial, so it has no ZECSs in the open boundary system, as we indicate again later in the discussion in Sec.\ref{sec4};at $\beta=1.1$, the energy band intersects the dotted line and the inner bandgap ESs appears, and the system behaves as topologically nontrivial, but the BSs and bulk are mixed at the ends of the $q_y$-axis, unlike research considering only the NN coupling \cite{A41};$\beta=1.2$, the energy band still intersects the dotted line, but for the first time the full bandgap inner ESs appears, as illustrated by the local enlarged map on the energy band, and again by the calculation of the open boundary system and the calculation of the nonlinear characteristic equation in Appendix \ref{smsec4};$\beta=1.3$, the energy band maintains the inner ESs of the band gap while the bulk band no longer intersects the dotted line for the first time in the topologically nontrivial case, implying that the corner states appears between 1.2 and 1.3, which is consistent with what is predicted by the periodic structure energy band derived earlier using the rigorous lattice sums technique, the exact value of which is discussed in detail later in Sec.\ref{sec4}.Under this parameter we also plot the dipole moment distribution around $q_y=-0.5\pi/d$, as in Figs.\ref{fig2}(e)-(g), where the z-axis denotes the real part of the normalized absolute value of the dipole moment, ES1 and ES2 denote the ESs above and below the gray dotted line, respectively, and BS denotes a bulk state.

\section{\label{sec4}ROBUST CORNER STATES AND BULK POLARIZATION}
In this section, we investigate in detail the topological properties of the Z-modes of the structure under open boundary conditions and compare them with the topological properties presented by the system under periodic and mixed boundary conditions.Specifically, we construct a structure with an open boundary of 20 UCs in both x- and y-directions, and similarly consider the coupling of each lattice point to all lattice points except itself.Under the linearization assumption, the coupling intensity is only related to the relative distance between the two dipoles, so even considering the fully coupled system still maintains the same $\mathrm{C}_4$ symmetry as the standard 2D SSH model.It can also be seen from Fig.\ref{fig1} that the second and third bands of the fully coupled band structure are always degenerate as $\beta$ keeps changing, which also indicates that the system is with $\mathrm{C}_4$ symmetry.In square lattice arrays with $\mathrm{C}_4$ symmetry, in order to make the zero-energy type-I corner states appear, the degeneracy of the energy band near the zero-energy is generally opened by breaking the TRS or $\mathrm{C}_4$ symmetry in past studies, which requires additional magnetic fields or modulation of the lattice geometry.Both of these methods are aimed at the emergence of zero-energy band gaps(ZEBGs). By analogy with this idea, it is natural to find a third way to realize ZEBGs in the square lattice, i.e., to move the degenerate band near the zero-energy without breaking the degeneracy and thus realize ZEBGs.At the same time, the practical optical system does not always strictly adhere to the assumption that only the NN coupling is present in the tight-binding model; non-zero non-NN coupling can shift the bands slightly but has little effect on the characteristics of the bands structure \cite{A36,A41}.The dipole arrays studied in this paper naturally have slowly decaying far-field interactions, and beyond the quasi-static approximation we consider its all dipole-dipole interactions, a factor that enables the band structures to be shifted, i.e., a third way to achieve zero-energy type-I corner states in a square lattice.
\begin{figure}[!htbp]
\centering
\includegraphics[width=0.90\linewidth]{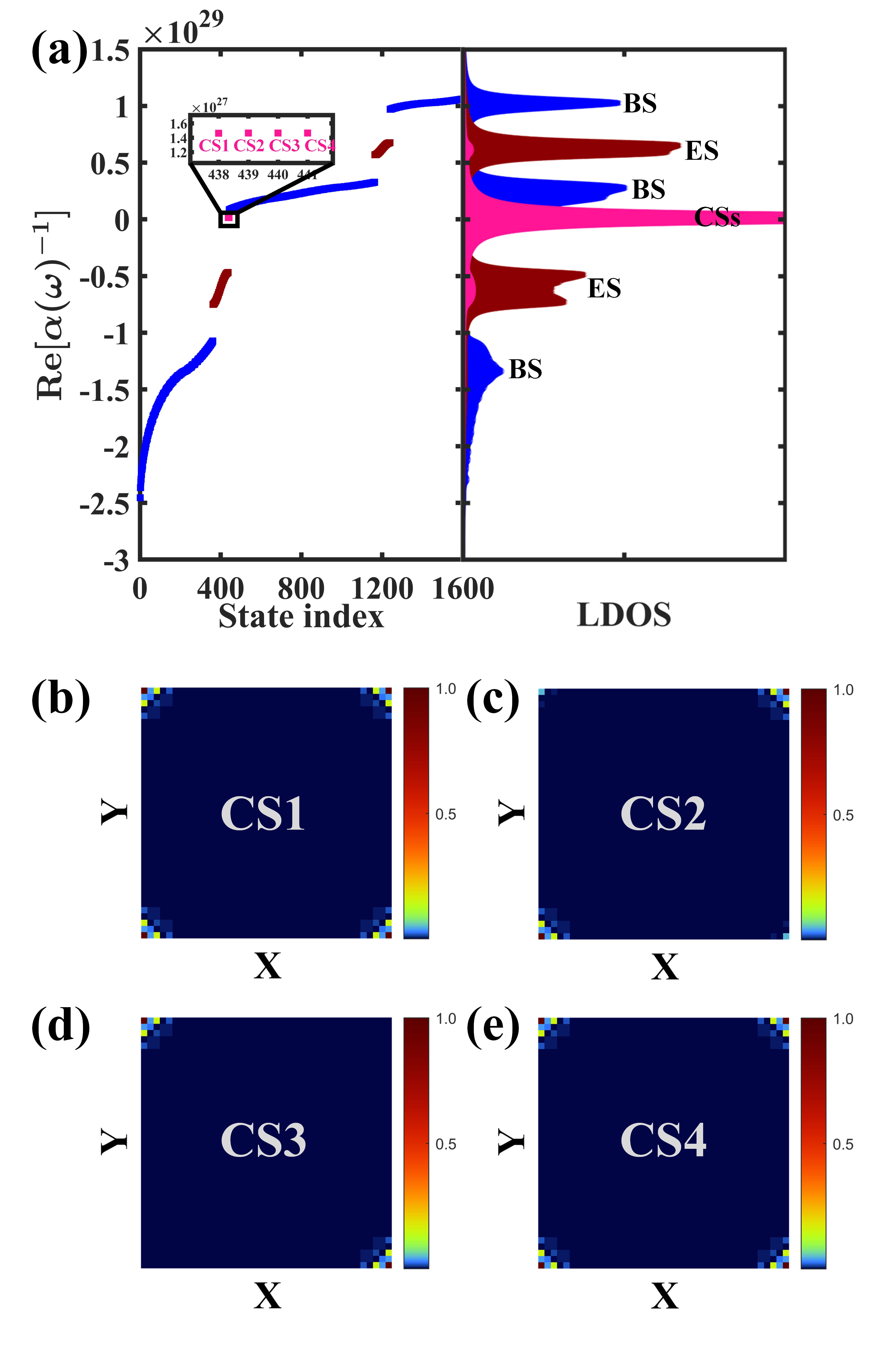}
\caption{Left panel of (a),the energy spectrum of the $\beta=1.3$ case with open boundary structure for 20 UCs in both x- and y-directions, blue represents the BSs, dark red are the ESs, deep pink are the CSs, and the enlarged view shows the degenerate CSs, four in total from CS1 to CS4.Right panel of (a) shows the local density of states(LDOS) corresponding to the energy spectrum, and the LDOS of the bulk, edge, and corner states are calculated by taking the mode components at the corresponding positions of the open boundary structure respectively, with the same color settings as the energy spectrum.(b)-(e) are heat maps of the  ZECSs at the enlarged view of the energy spectrum, showing the distribution of the absolute values of the corresponding dipole moments over the open boundary structures.}
\label{fig3}
\end{figure}

\begin{figure}[!htbp]
\centering
\includegraphics[width=0.95\linewidth]{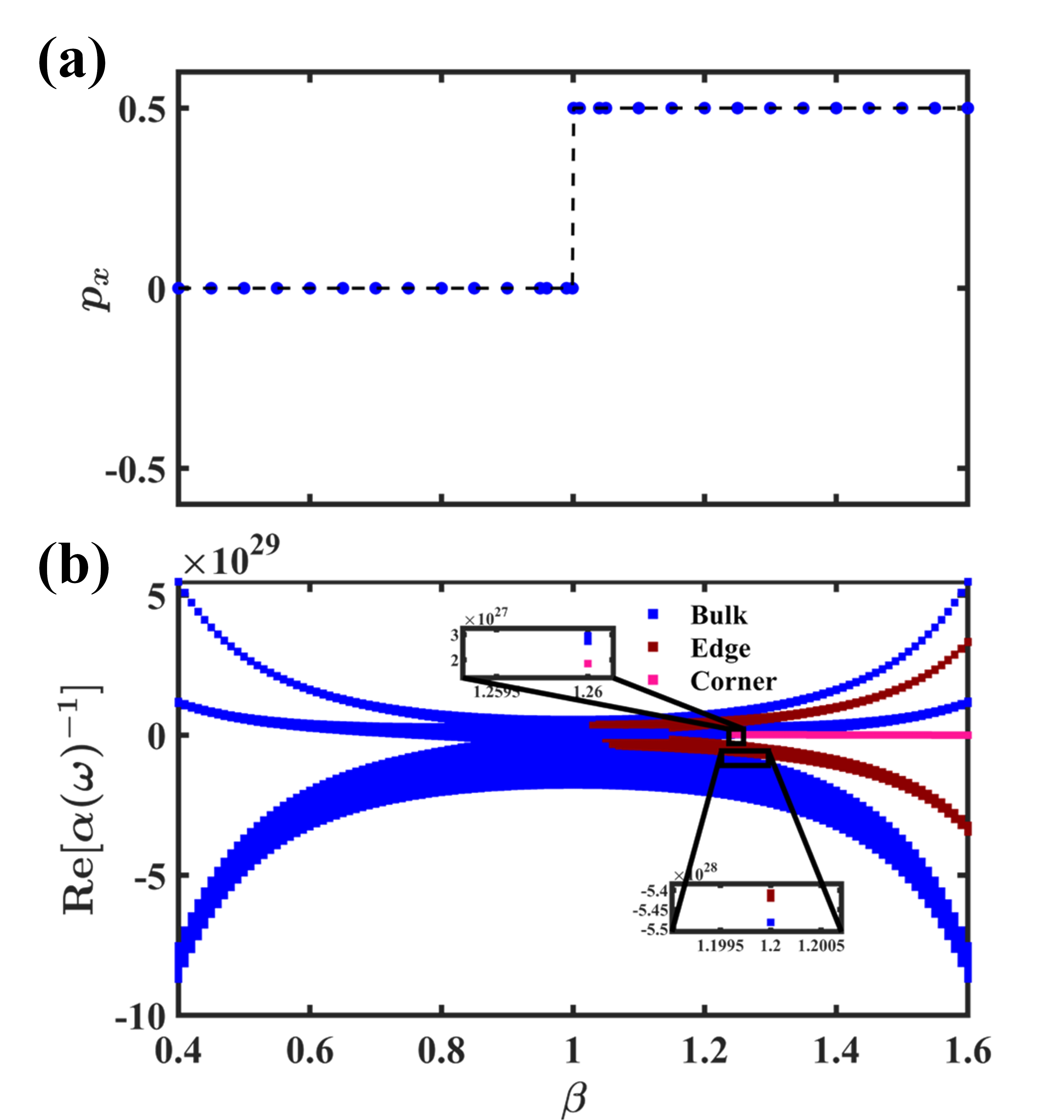}
\caption{The bulk polarization and open boundary energy spectrum in the range of $\beta$ from 0.4 to 1.6, with dipole approximation holds.(a) The bulk polarization of the fourth band from bottom to top in the bulk energy spectrum ,and is quantized to 0 or 0.5 in all $\beta$ ranges, with $\beta$ = 1 for a phase transition point.(b) Open boundary energy spectrum (in steps of 0.01) for different values of $\beta$, blue for the BSs, dark red for the ESs, and deep pink for the CSs, when $\beta \ge 1.2$, the stable off-bulk ESs appears, and $\beta \ge 1.26$ CSs are separated from the bulk.}
\label{fig4}
\end{figure}

The structure of size $20\times20$ considered in this section it has an eigenmatrix of size $1600\times1600$, and since it is not a periodic structure, resulting in each term of the eigenmatrix not being an infinite series, the lattice sums technique does not apply here.The simplified DGF in Z modes is: $\mathbf{G}=\frac{e^{-i k R}}{4 \pi R}\left\{1-\frac{i k R+1}{k^2 R^2}\right\}$, and combining it with Eq.(\ref{eq1}), using the direct summation  we can compute each element in the eigenmatrix.Taking $\beta=1.3$, the band structures are shown in the left panel of Fig.\ref{fig3}(a).As in Sec.\ref{sec2}, $1/\alpha$ is taken as the eigenvalue here and only its real part is plotted in order to simplify the calculation.For rigorous considerations, the real and imaginary parts of the finite structure energy bands with $\omega$ as the eigenvalues are plotted in Appendix \ref{smsec5}. In reality, the topological properties presented by the energy bands in these two cases are identical except for the difference in the up-and-down flip of the energy bands, while the imaginary part of the energy bands tends to be three orders of magnitude smaller than the real part.Strictly speaking, due to the complex coupling coefficients the system considered in this paper is a Non-Hermitian system \cite{A55}, whose most characteristic feature is the failure of the so-called bulk-edge correspondence \cite{B2}, i.e., the difference between the topological properties predicted by the periodic structure and those under open boundary conditions, but the Non-Hermitian system in this paper still maintains the bulk-edge correspondence.As shown in Fig.\ref{fig1}(e) when $\beta$=1.3 the periodic structure energy bands are shifted upwards and the ZEBGs appears, under the open boundary condition when $\beta$=1.3 the energy band structure shown in Fig.\ref{fig3}(a) also has states near zero energy, and the periodic structure accurately predicts the topological states under the open boundary.This is further illustrated in the calculation of the bulk polarization and the drawing of the band structures for a large range of $\beta$  in Fig.\ref{fig4} later.

The right panel of Fig.\ref{fig3}(a) is the so-called local density of states (LDOS) \cite{A24}, which is calculated by considering all eigenvalues and corresponding modes.As with the calculation of DOS in Sec.\ref{sec2}, only the right vector and the corresponding eigenvalues are considered here, while the small quantities are taken to be three orders of magnitude smaller than the eigenvalues.Unlike the DOS calculation where LDOS needs to be specified at certain sites \cite{A25}, here we have selected the components of the modes in the bulk, boundary and corners of the structure to calculate LDOS and plotted them together in the right panel of Fig.\ref{fig3}(a),the parameters in the right panel take the same parameters as in the left panel of Fig.\ref{fig3}(a), which accurately distinguishes the bulk states (BSs), edge states (ESs) and corner states (CSs) of the  bands.Meanwhile Fig.\ref{fig3}(b)-(e) plots the degenerate ZECSs at Fig.\ref{fig3}(a).
The enlarged view on the energy spectrum shows a total of four modes from CS1 to CS4,and the heat map is drawn to show the distribution of the absolute values of the corresponding dipole moments on the open boundary structure, it can be seen that the vast majority of the energy is concentrated on the single SiC nanoparticles in the corners, which is  more better than the localization of the type-II and type-III corner states separated from the ESs.

Under the premise that the dipole approximation \cite{A39} holds, the variation of the bulk polarization (a) and the open boundary bands structure (b) are mapped for all cases of $\beta$ taking values as in Fig.\ref{fig4}.Since the energy band structures with the eigenvalue of $1/\alpha$ is geometrically flipped $180^{\circ}$ up and down with respect to the standard energy band structures with the eigenvalue of $\omega$, Fig.\ref{fig4}(a) is plotted the bulk polarizations of the fourth energy band, which is relative to the bulk polarization of the first energy band in the standard energy band structures.In this section, only the bulk polarization of isolated bands is calculated,in Appendix \ref{smsec2}, the Wannier bands of the degenerate  bands formed by the second and third  bands are also calculated.Combined with Fig.\ref{fig4}(b), it can be seen that the quantized bulk polarization  accurately captures the topological phase transition of the system.Since the system in this paper has kept the ${C}_4$ symmetry, which leads to equal bulk polarization in the x-direction and y-direction ($p_x=p_y$)\cite{A28,A29,A36}, only the calculation of $p_x$ is considered here, and the Wilson loop \cite{A34,A35} is used here to calculate the bulk polarization for the stability of the numerical calculation.

For isolated bands, the Wilson loop is a complex number and the Wannier center ${W}$ evolves with $q_y$ as:
\begin{align}
{W}\left(q_y\right)=\frac{1}{2 \pi} \mathrm{Im}\left\{\mathrm{In}\left[\prod_{i=1}^{N-1} \frac{\langle p_{i,n}^z \mid p_{i+1,n}^z\rangle}{\left|\langle p_{i, n}^z \mid p_{i+1, n}^z\rangle\right|}\right]\right\},
\label{eq9}
\end{align}
$p_x$ is the average of the evolution of the Wannier center with $q_y $ in the first Brillouin zone ($q_x,q_y\in[-\pi/d,\pi/d]$),the meanings of the parameters of Eq.(\ref{eq9}) are: $n$ denotes the $n$th band, $(N-1) \Delta q_x=2\pi/d$,$p_{i, n}^z=p_n^z[-\pi/d+(i-1) \Delta q_x, q_y]$, where the superscript $ z$ denotes the $ z$ component of the dipole moment.Also there is the gauge: $p_{1, n}^z=p_{N,n}^z$.

For the degenerate bands, the Wilson loop is a matrix, and assuming only the degenerate bands formed by the $n$, $n+1$th  band are considered, the evolution of the Wannier center ${W}$ with $q_y$ is given by:
\begin{align}
W\left(q_y\right)=\frac{1}{2 \pi} \operatorname{Im}\left\{\operatorname{In}\left[\operatorname{eig}\left(\prod_{i=1}^{N-1} U_{i, i+1}\right)\right]\right\},
\label{eq10}
\end{align}
where '$\mathrm{eig}$' denotes the eigenvalue of the matrix and $U_{i,i+1}$ is a matrix:
\begin{align}
U_{i, i+1}=\left(\begin{array}{cc}
\langle p_{i, n}^z \mid p_{i+1, n}^z\rangle & \langle p_{i, n}^z \mid p_{i+1, n+1}^z\rangle \\
\langle p_{i, n+1}^z \mid p_{i+1, n}^z\rangle & \langle p_{i, n+1}^z \mid p_{i+1, n+1}^z\rangle
\end{array}\right),
\label{eq11}
\end{align}
the meaning of $ p_{i, n}^z$ is the same as the calculation of the Wannier center of the isolated bands, and the rest parameters mean the same as those of the isolated bands. It should also be noted that the dipole moments in the above equations are all right vectors,this consideration is reasonable according to Ref.\cite {A34}.According to the bulk-edge correspondence, the presence of non-zero bulk polarization protects the presence of ESs in the corresponding direction\cite{A28}, while the presence of CSs is related to the non-zero edge polarization\cite{A41}.Since there is only one band under the band gap considered in this work, and also the presence of CSs is related to the non-zero bulk polarization due to the $\mathrm C_{4}$ symmetry\cite{A36,A41}, it is not necessary to calculate the edge polarization here.

Fig.\ref{fig4}(b) with $\beta$ varying in steps of 0.01, and unlike studies considering only NN coupling \cite{A41}, the stable ESs of the discrete bulk does not appear immediately when $\beta>$1.Between 1 and 1.2 the mode numbers and positions of the ESs are unstable because they are mixed with the bulk, and only when $\beta \ge1.2$ the ESs are completely separated from the bulk thus showing stable mode numbers and positions, which is consistent with the energy spectrum of the Fig.\ref{fig2}(c).Also between 1 and 1.26 the CSs and the bulk are mixed together and exhibit the bound states in the continuum (BIC)\cite{A01,A02,A03}, which is not labeled in Fig.\ref{fig4}(b) due to the instability.When $\beta \ge1.26$, the CSs is separated from the bulk and behaves as a stable ZECSs, which is consistent with the range of CSs appearance predicted by the energy spectrum of Fig.\ref{fig1} periodic structure and Fig.\ref{fig2} nanoribbons, which also proves side-by-side the correctness of the lattice sums technique used earlier.Also Appendix \ref{smsec4} investigates the topological properties of the open boundary system in the variable $\omega$ case, which is consistent with what is described in this section and supports the correctness of the linearization assumption in this paper.

\section{\label{sec5}DISCUSSION AND CONCLUSIONS}
In summary, we investigate square lattice topological insulators beyond the NN coupling assumption, surpassing the quasi-static approximation in previous topological research on dipole arrays and extending the study of full coupling in 1D SSH model dipole arrays \cite{A50} to fully coupled square lattice arrays in the 2D case.
The lattice sums technique derived in this work can be applied to different types of periodic 2D dipole arrays.
In contrast to the method of Ref \cite{A52}, the technique of this paper can be applied to the case of multiple lattice points within a unit cell, and the accuracy of this method is demonstrated by a detailed comparison with direct summation over a large parameter range in Appendix \ref{smsec1}.
Although the topological properties of the nanoribbons are studied using the truncated summation method, the lattice sums technique considering the half-open and half-periodic boundary conditions is still derived in Appendix \ref{smsec1} and the projection bands are plotted within its validity, which present the same topological properties as the truncated summation.Moreover, the topological properties of the bulk bands drawn based on the lattice sums technique, the projection bands drawn based on the truncated summation and the open boundary energy spectrum drawn based on the direct summation are identical, which also shows the effectiveness of the truncated summation.Since it is beyond the scope and research topic of this paper, the failure causes of the lattice sums technique under mixed boundary conditions are not studied in detail in this work.

Beyond the conventional 2D SSH model, it is possible to achieve HOTPs without breaking the $\mathrm C_{4}$ symmetry and TRS of the square lattice arrays while considering long-range coupling, as illustrated by recent similar research in acoustics \cite{A27} and circuits \cite{A26}.Compared to them this work further considers all coupling interactions, in this case the bulk polarization still accurately captures the topological properties of the system and is found to support robust ZECSs in the square lattice dipole arrays based on the 2D SSH model, which is not present in similar studies considering only the NN coupling \cite{A41}.

The linearization assumption is applied in the main text, and the topological properties of the open boundary system in the variable $\omega$ case are again studied in Appendix \ref{smsec4}, which is the same as described in the main text and illustrates the correctness of the linearization assumption.The combination of the CDM and the lattice sums technique accurately captures the topological properties of the system and provides a relatively accurate theoretical tool for the experimental investigations of the topology in dipole arrays, while providing new insights into the effect of long-range interactions on the topology.

\begin{acknowledgments}
Many thanks to Dr. Simon R Pocock of Imperial College London, Mr. Ivica Stevanović of Swiss Federal Office of Communications and Dr. Langland Xiong of Fudan University for their selfless help and enlightening discussions.And this work was supported by the National Natural Science Foundation of China (Grant numbers 61865009, 61927813).
\end{acknowledgments}


\appendix
\newpage
\begin{widetext}
\section{\label{smsec1}EFFECTIVENESS OF THE LATTICE SUMS TECHNIQUE}
In this section, the lattice sums technique for periodic and mixed boundary conditions are derived in detail and their accuracy is verified separately. The lattice sums technique for mixed boundary conditions is more complicated and has its own conditions of use, which are explained in detail here.
\subsection{Periodic boundary conditions}
As mentioned in the main text, most of the elements in the eigenmatrix under the periodic boundary condition are repeated, so only the following need to be discussed here: $ \Sigma \mathbf{G}[\begin{array}{l}\vec{R}+\vec{r}_{u v}\end{array}] \cdot t,\Sigma^{\prime} \mathbf{G}(\vec{R}) \cdot t $, and the specific meaning of each parameter is given in the main text.

First discuss the Ewald series form of $\Sigma \mathbf{G}[\begin{array}{l}\vec{R}+\vec{r}_{u v}\end{array}] \cdot t$, refer to Eq.(\ref{eq2}) in the main text which can be written as:

\begin{align}\label{smeq1}
&\Sigma \mathbf G\left[\vec{R}+\vec{r}_{u v}\right] \cdot t \notag\\
&=\sum_{(m, n) \in Z}\left\{\hat{\mathbf I}+\frac{1}{k^{2}} \nabla \otimes \nabla\right\} \frac{e^{-i k\left|\vec{R}+\vec{r}_{u v}\right|}}{4 \pi\left|\vec{R}+\vec{r}_{u v}\right|} \cdot e^{i \vec{q} \cdot \vec{R}} \\
&=\sum_{(m, n) \in Z}\left\{\hat{\mathbf I} \cdot \frac{e^{-i k\left|\vec{R}+\vec{r}_{u v}\right|}}{4 \pi\left|\vec{R}+\vec{r}_{u v}\right|} \cdot e^{i \vec{q} \cdot \vec{R}}+\frac{1}{k^{2}} \nabla \otimes \nabla\left(\frac{e^{-i k\left|\vec{R}+\vec{r}_{u v}\right|}}{4 \pi\left|\vec{R}+\vec{r}_{u v}\right|} \cdot e^{i \vec{q} \cdot \vec{R}}\right)\right\}\notag,
\end{align}
the meanings of the parameters in the above equation are the same as in the main text. And note the final result in Eq.(\ref{smeq1}), which is divided into a scalar part and a dyadic part.

Discussing first the scalar part, under the Ewald method the slowly converging Green's function is transformed into a combination of modified real space ($G_R$) and k-space ($G_K$) levels with Gaussian convergence rate \cite{A10}:
\begin{align}\label{smeq2}
&\sum_{(m, n) \in Z} \frac{e^{-i k\left|\vec{R}+\vec{r}_{u v}\right|}}{4 \pi\left|\vec{R}+\vec{r}_{u v}\right|} e^{i \vec{q} \cdot \vec{R}} \notag\\
&=\sum_{(m, n) \in Z} \frac{e^{-i k \cdot R r_{m n}}}{4 \pi R r_{m n}} \cdot e^{i \vec{q} \cdot \vec{R}}\notag\\
&=G_R+G_K\\
&=\frac{1}{8 \pi} \sum_{(m, n) \in Z} \frac{e^{i \vec{q} \cdot \vec{R}}}{R r_{m n}}\left\{\sum_{\pm} e^{\pm i k \cdot R r_{m n}} \operatorname{erfc}\left(R r_{m n} \cdot E \pm \frac{i k}{2 E}\right)\right\}_{\text{R}}+\notag\\
&~~~~\frac{1}{4 d^2}\sum_{(m, n) \in Z} e^{i K_{X} \cdot r_{x}+i K_{Y} \cdot r_{y}} \cdot \frac{1}{\Gamma_{m n}}\left\{\sum_{\pm} e^{\pm r_{z}\cdot \Gamma_{m n}} \operatorname{erfc}\left(\frac{\Gamma_{m n}}{2 E} \pm r_{z} E\right)\right\}_{\text{K}},\notag
\end{align}
where $Rr_{m n}=|\vec{R}+\vec{r}_{u v}|$ denotes the distance between any two lattice points in the periodic structure, $\vec{R}=m d \cdot \hat{x}+n d \cdot \hat{y}$ denotes the vector connecting any two UCs, $m,n$ are the summation parameters, which take any integer value in Eq.(\ref{smeq2}), and $d$ is the lattice constant.$\vec{r}_{u v}=\vec{r}_u-\vec{r}_v=r_{x} \cdot \hat{x}+r_{y} \cdot \hat{y}+r_{z} \cdot \hat{z}$ ($u,v$ are the numbering of particles in the same unit cell, see Fig.1(a) in the main text for details.) denotes the vector connecting any two different lattice points within the same unit cell, and its specific values are given in Sec.II of the main text.Since the derivation of Eq.(\ref{eq2}) in the main text requires the z-component of the distance vector, the derivation of Eq.(\ref{smeq2}) here also requires the consideration of $r_z$, but this research considers a 2D array, so $r_z$ is set to zero in the final calculation.$\vec{q}=q_{x} \cdot \hat{x}+q_{y} \cdot \hat{y}$ denotes the Bloch wave vector as in the main text, and they take values in the range of the reduced Brillouin zone along the high symmetry points. $\Gamma_{m n}^{2}=K_{X}^{2}+K_{Y}^{2}-k^{2}$,$k$ is the wave vector when the relative permittivity of the background is set to 1, $K_{X}=2\pi\cdot m/d-q_{x}$, $K_{Y}=2\pi\cdot n/d-q_{y}$. The $E$ is the splitting parameter and the specific expression is:
\begin{align}\label{smeq3}
E=\max \left(\sqrt{\frac{\pi}{d^2}}, \frac{\mathrm{Re}(k)}{2 H}\right),
\end{align}
where $H$ is an empirical parameter, set here to 3 \cite{A11,A08},$\mathrm{erfc}(\cdot)$ is the complex complementary error function \cite{A08},its quick calculation code can be found in Ref.\cite{web}.

Then discussing the dyadic part:
\begin{align}\label{smeq4}
\begin{split} 
&\sum_{(m, n) \in Z} \frac{1}{k^{2}} \nabla \otimes \nabla\left(\frac{e^{-i k\left|\vec{R}+\vec{r}_{u v}\right|}}{\left.4 \pi |\vec{R}+\vec{r}_{u v}\right.|} \cdot e^{i \vec{q} \cdot \vec{R}}\right) \\
&=\frac{1}{k^{2}}\left\{\nabla \otimes \nabla G_{K}+\nabla \otimes \nabla G_{R}\right\},
\end{split}
\end{align}
it can be seen that Eq.(\ref{smeq4}) is the sum of the dyadic series in real-space $\nabla \otimes \nabla G_{R}$ and the dyadic series in k-space $\nabla \otimes \nabla G_{K}$.The specific expression for $\nabla \otimes \nabla G_{R}$ is \cite{A12,A05}:
\begin{align}\label{smeq5}
\begin{split}
&\frac{1}{k^2}\nabla \otimes\nabla G_{R} \\
&=\frac{1}{k^2}\cdot\frac{1}{8 \pi} \sum_{(m, n) \in Z} e^{i \vec{q} \cdot \vec{R}} \cdot \hat{\mathbf{F}}_{3 \times 3},
\end{split}
\end{align}
which:
\begin{align}\label{smeq6}
\begin{split}
\hat{\mathbf{F}}_{3\times 3}=\left(\frac{\partial f}{\partial R r_{m n}} \cdot \frac{1}{R r_{m n}^{2}}-\frac{f}{R r_{m n}^{3}}\right) \cdot\left[\begin{array}{ccc}
1 & 0 & 0 \\
0 & 1 & 0 \\
0 & 0 & 1
\end{array}\right]+~~~~~~~~~~~~~~~~~~~~~~~\\
\left(\frac{\partial^{2} f}{\partial R r_{m n}^{2}} \cdot \frac{1}{R r_{m n}}-3 \frac{\partial f}{\partial R r_{m n}} \cdot \frac{1}{R r_{m n}^{2}}+\frac{3 f}{R r_{m n}^{3}}\right) \frac{\left(\vec{R}+\vec{r}_{u v}\right) \otimes\left(\vec{R}+\vec{r}_{u v}\right)}{R r_{m n} \cdot R r_{m n}},
\end{split}
\end{align}
assume here that:$\large f\left(R r_{m n}\right)= e^{i k \cdot R r_{m n}} \cdot \operatorname{erfc}\left(R r_{m n} E+\frac{i k}{2 E}\right)+
 e^{-i k \cdot R r_{m n}} \cdot \operatorname{erfc}\left(R r_{m n} E-\frac{i k}{2 E}\right)$ and the first order derivative of $\large f$ with respect to $\large Rr_{mn}$ is:
 \begin{equation}\label{smeq7}
 \begin{split}
\frac{\partial f}{\partial R r_{m n}}=i k\left[e^{i k \cdot R r_{m n}} \operatorname{erfc}\left(\beta_{+}\right)-e^{-i k \cdot R r_{m n}} \operatorname{erfc}\left(\beta_{-}\right)\right]-(A+B),
\end{split}
\end{equation}
which:$ \large A=\frac{2 E}{\sqrt{\pi}} \cdot e^{-\left(\beta_{-}\right)^2} \cdot e^{-i k \cdot R r_{m n}}$, $\large B=\frac{2 E}{\sqrt{\pi}} \cdot e^{-\left(\beta_{+}\right)^2} \cdot e^{i k \cdot R r_{m n}}$;$\large \beta_{+}=R r_{m n} E+\frac{i k}{2 E}$, $\large \beta_{-}=R r_{m n} E-\frac{i k}{2 E}$.The second order derivative of $f$ with respect to $Rr_{mn}$ is:
\begin{align}\label{smeq8}
\frac{\partial^2 f}{\partial R r_{m n}^2}=-k^2 f(Rr_{mn})+2 i k(A-B)+C+D,
\end{align}
which: $C=2 E  \beta_{-} \cdot A$, $D=2 E  \beta_{+} \cdot B$. The third order derivative of $f$ with respect to $Rr_{mn}$ is:
\begin{align}\label{smeq9}
\begin{split}
\frac{\partial^3 f}{\partial R r_{m n}^3}
=&-k^2 \frac{\partial f}{\partial R r_{m n}}+3 i k(D-C)+\left(2 k^2+2 E^2\right)(A+B)-2 E\left(\beta_{-} \cdot C+\beta_{+} \cdot D\right),
\end{split}
\end{align}
when $Rr_{mn}$ = 0 there are:
\begin{align}\label{smeq10}
\begin{split}
&f\left(R r_{m n}=0\right)=2, \quad f^{\prime}\left(R r_{m n}=0\right)=-\frac{4 E}{\sqrt{\pi}} e^{\frac{k^2}{4 E^2}}+i k\left[\operatorname{erfc}\left(\frac{i k}{2 E}\right)-\operatorname{erfc}\left(-\frac{i k}{2 E}\right)\right] \\
&f^{\prime \prime}\left(R r_{m n}=0\right)=-2 k^2, \quad f^{\prime \prime \prime}\left(R r_{m n}=0\right)=\frac{8 E^3}{\sqrt{\pi}} e^{\frac{k^2}{4 E^2}}-k^2 f^{\prime}\left(R r_{m n}=0\right),
\end{split}
\end{align}
note that the superscript ${}^{\prime}$ there indicates the derivative for $Rr_{mn}$. Eq.(\ref{smeq10}) will be used later in the discussion.Returning to Eq.(\ref{smeq5}), since this work is considering the Z-modes, it is only necessary to take the (3,3) term of the $\hat{\mathbf{F}}_{3\times 3}$ to obtain the $\nabla \otimes \nabla G_{R}$ expression in the Z-modes as follows:
\begin{align}\label{smeq11}
\begin{split}
&\left.\frac{1}{k^2} \nabla \otimes \nabla G_R\right |_z \\
&=\frac{1}{k^2}\cdot \frac{1}{8 \pi} \sum_{(m, n) \in Z} e^{i \vec{q} \cdot \vec{R}} \cdot\left(\frac{\partial f}{\partial R r_{m n}} \cdot \frac{1}{R r_{m n}^2}-\frac{f}{R r_{m n}^3}\right),
\end{split}
\end{align}
Eq.(\ref{smeq11}) is one of the formulas that will eventually be used in the calculation.Then consider $ \nabla \otimes \nabla G_ {K} $, its specific form is:
\begin{align}\label{smeq12}
\begin{split}
&\frac{1}{k^2}\nabla \otimes \nabla G_K\\
&=\frac{1}{k^2}\begin{pmatrix}
 i K_X\cdot iK_X G_K &i K_X\cdot i K_Y G_K  &i K_X \cdot \nabla G_K^{z} \\
 i K_Y \cdot iK_X G_K &i K_Y\cdot iK_Y G_K  &i K_Y \cdot \nabla G_K^{z} \\
 i K_X \cdot \nabla G_K^{z} & i K_Y \cdot \nabla G_K^{z} & \partial_z \nabla G_K^{z} 
\end{pmatrix},
\end{split}
\end{align}
the $\nabla G_K^{z}$ denotes the z-component of the $\nabla G_K$ vector. Since the interest in this work is on the Z-modes, only $\partial_z \nabla G_K^{z}$ is considered here:
\begin{align}\label{smeq13}
\begin{split}
&\frac{1}{k^2}\cdot\partial_{z} \nabla G_{K}^{z}\\
&=\frac{1}{k^2}\frac{1}{4 d^2} \sum_{(m,n)\in Z} e^{i K_{X} \cdot r_{x}+i K_{Y}\cdot r_{y}} \times \\
&\left\{\sum_{\pm} \Gamma_{m n} \cdot e^{\pm \Gamma_{m n}\cdot r_{z}} \mathrm{erfc}\left(\frac{\Gamma_{mn}}{2 E} \pm r_{z} E\right)-\right.
\left.\frac{4 E}{\sqrt{\pi}} e^{-\frac{\Gamma_{m n}^{2}}{4 E^{2}}-r_{z}^{2} E^{2}}\right\},
\end{split}
\end{align}
Eq.(\ref{smeq13}) is one of the formulas that will be used for the final calculation. The sum of Eq.(\ref{smeq11}) and Eq.(\ref{smeq13}) is the specific form of Eq.(\ref{smeq4}) in the Z-modes. Combined with Eqs.(\ref{smeq1}), (\ref{smeq2}), (\ref{smeq4}) is the specific form of the Ewald series of $\Sigma \mathbf{G}[\begin{array}{l}\vec{R}+\vec{r}_{u v}\end{array}] \cdot t$ in the Z-modes, and they only need to consider a few terms to reach convergence.

Then discussing the Ewald series of $\Sigma^{\prime} \mathbf{G}(\vec{R}) \cdot t$, refer to Eq.(\ref{eq2}) of the main text it can be written as:
\begin{align}\label{smeq14}
\begin{split}
&\Sigma^{\prime} \mathbf G(\vec{R}) \cdot t\\
&=\sum_{(m, n) \neq(0,0)}\left\{\hat{\mathbf I}+\frac{1}{k^{2}} \nabla \otimes \nabla\right\} \frac{e^{-i k|\vec{R}|}}{4 \pi|\vec{R}|} \cdot e^{i \vec{q} \cdot \vec{R}}\\
&=\sum_{(m, n) \neq(0,0)}\left\{\hat{\mathbf I} \cdot \frac{e^{-i k|\vec{R}|}}{4 \pi|\vec{R}|} \cdot e^{i \vec{q} \cdot \vec{R}}+\frac{1}{k^{2}} \nabla \otimes \nabla\left(\frac{e^{-i k|\vec{R}|}}{4 \pi|\vec{R}|} \cdot e^{i \vec{q} \cdot \vec{R}}\right)\right\},
\end{split}
\end{align}
the specific form of Eq.(\ref{smeq14}) depends on the specific form of the Eq.(\ref{smeq1}), which also needs to be divided into a scalar part and a dyadic part to discuss. 

The scalar part is discussed first as follows:
\begin{align}\label{smeq15}
\begin{split}
& \sum_{(m, n) \neq 0} \frac{e^{-i k|\vec{R}|}}{4 \pi|\vec{R}|} e^{i \vec{q} \cdot \vec{R}} \\
&= \lim _{\left|\vec{r}_{u v}\right| \rightarrow 0}\left\{\sum_{(m, n) \in Z}\left[\frac{e^{-i k\left|\vec{R}+\vec{r}_{u v}\right|}}{4 \pi\left|\vec{R}+\vec{r}_{u v}\right|} e^{i \vec{q} \cdot \vec{R}}\right]-\frac{e^{-i k\left|\vec{r}_{u v}\right|}}{4 \pi\left|\vec{r}_{u v}\right|}\right\} \\
&= \lim _{\left|\vec{r}_{u v}\right| \rightarrow 0}\left\{G_K+\left[G_R-\frac{e^{-i k\left|\vec{r}_{u v}\right|}}{4 \pi\left|\vec{r}_{u v}\right|}\right]\right\} \\
&\approx \left.G_K\right|_{\left(r_x, r_y\right)=(0,0)}+\left.G_{R,(m, n) \neq 0}\right|_{\left(r_x, r_y\right)=(0,0)}+\\ 
&~~~~~\frac{1}{4 \pi}\left\{i k \cdot \operatorname{erfc}\left(\frac{i k}{2 E}\right)-\frac{2 E}{\sqrt{\pi}} \cdot e^{\frac{k^2}{4 E^2}}\right\},
\end{split}
\end{align}
where $G_R$ and $G_K$ are the $G_R$ and $G_K$ in the Eq.(\ref{smeq2}). The derivation of Eq.(\ref{smeq15}) uses the so-called regularization method, the specific steps can be found in Ref.\cite{A05}.
\begin{figure}[!htbp]
\begin{minipage}{0.49\textwidth}
\centering
\includegraphics[width=8cm]{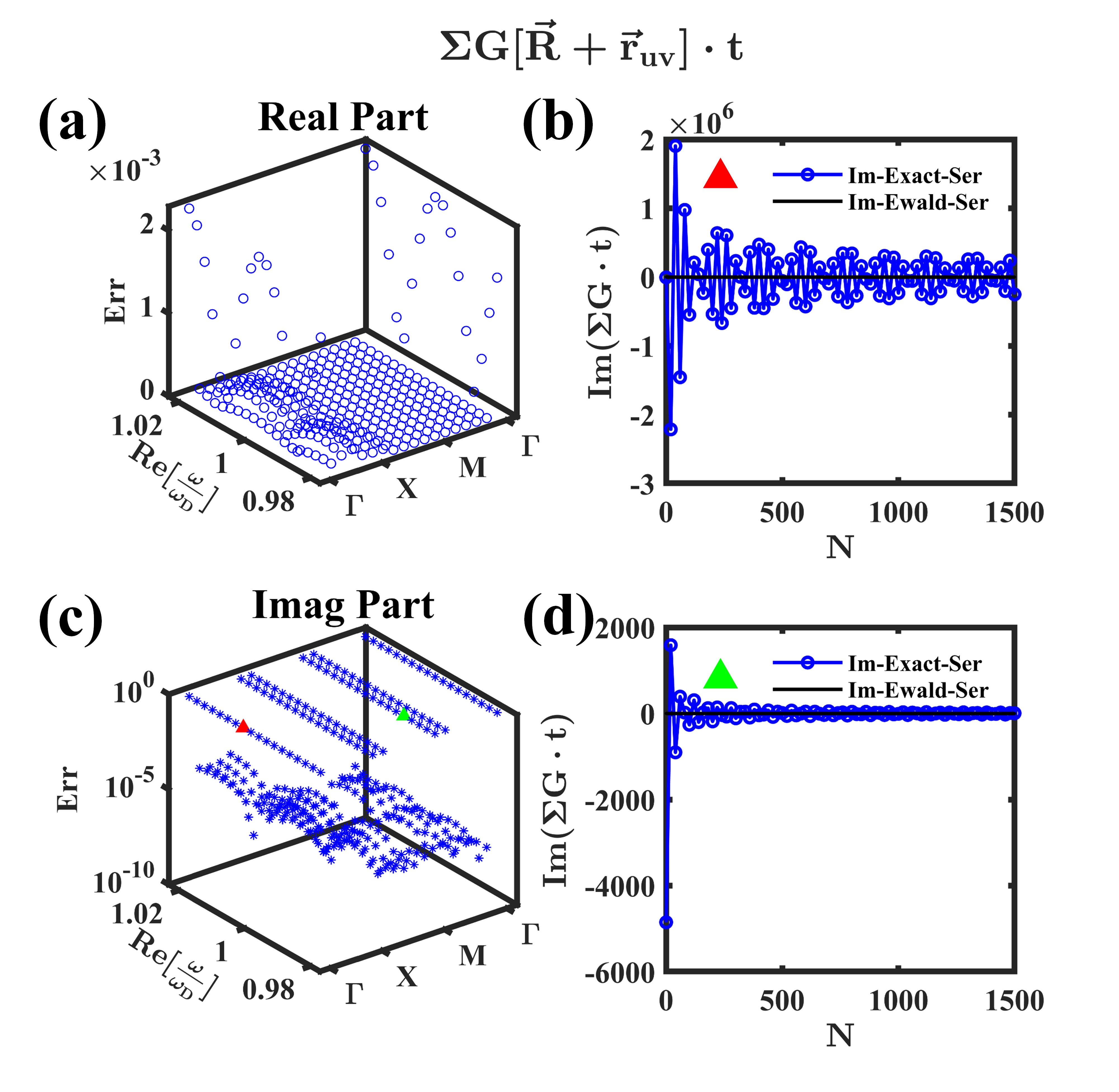}
\caption{\label{smfig1}Comparison of the direct summation of $\Sigma \mathbf{G}[\begin{array}{l}\vec{R}+\vec{r}_{u v}\end{array}] \cdot t$ (exact series) with the Ewald series.(a) Error (Err) between the real part of the Ewald series and the exact series, with the x-axis being the Brillouin zone path when plotting the periodic energy spectrum and the y-axis being the value of the real part of $\omega$ from 0.98$\omega_D$ to 1.02$\omega_D$, hawking all possible values of $\omega$ for the periodic energy spectrum.(b) The imaginary parts of the exact and Ewald series evolve as $\mathrm{N}(m,n=0,\pm1,... ,\pm \mathrm{N})$ varies, and since the Ewald series requires only a few terms to reach convergence, the $\mathrm{N}$ of it is fixed to 5. The black line denotes the Ewlad series, and the line marked by the blue small circles denotes the exact series. As $\mathrm{N}$ changes, the exact series keeps oscillating around the Ewald series and slowly converges to it.(c) The error (Err) between the imaginary part of the Ewald series and the exact series, the z-axis has log scale, and two points with large Err are marked randomly with red and green triangles respectively, the evolution of the series under the corresponding parameters are plotted in (b), (d), and the other settings are the same as in (a).(d) The same as (b) describes the evolution of the imaginary part of both series as the $\mathrm{N}$ varies.}
\end{minipage}
\hfill
\begin{minipage}{0.49\textwidth}
\centering
\includegraphics[width=7.45cm,height=7.8cm]{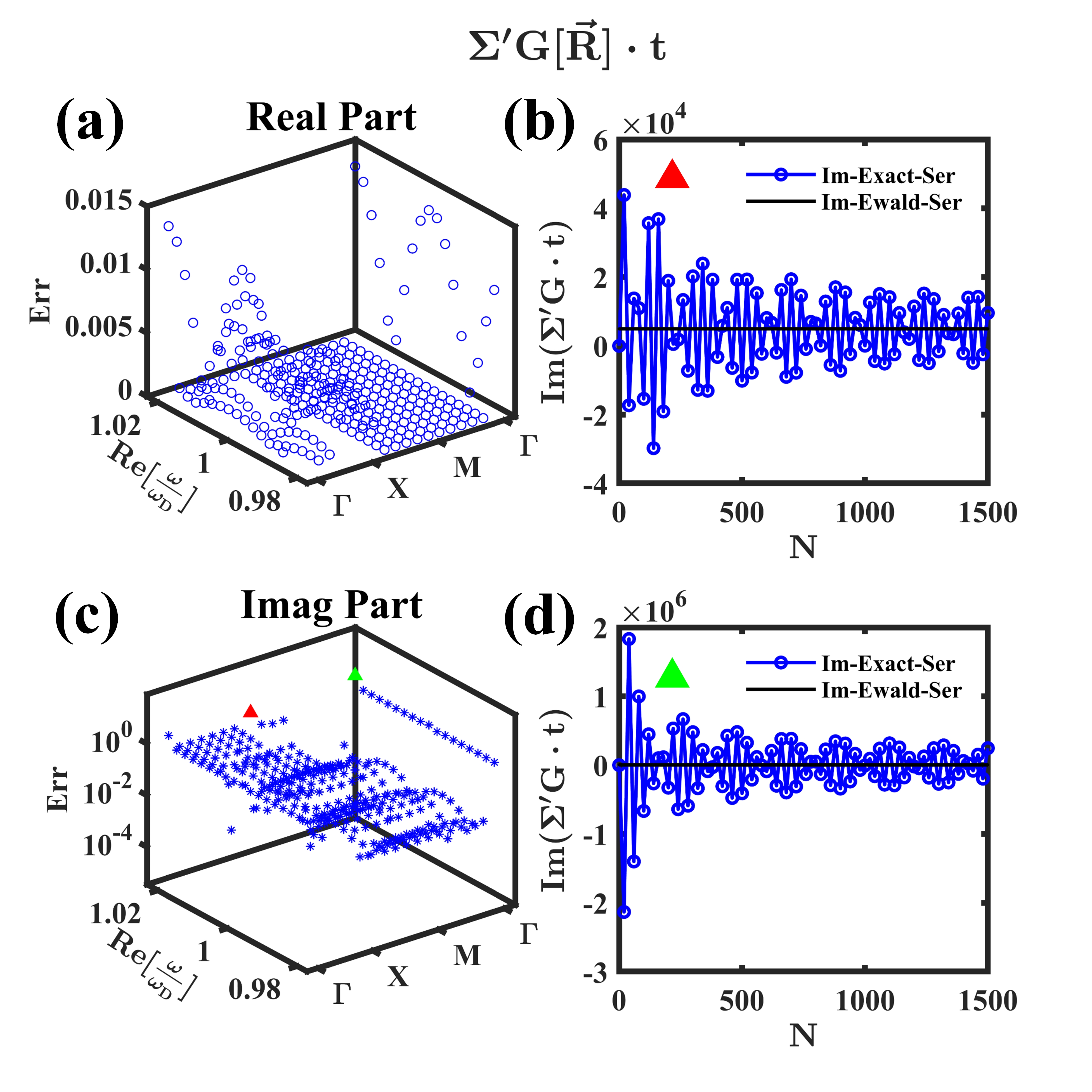}
\caption{\label{smfig2}Comparison of the direct summation of $\Sigma^{\prime} \mathbf{G}(\vec{R}) \cdot t$ (exact series) with the Ewald series.(a) Error (Err) between the real part of the Ewald series and the exact series, with the x-axis being the Brillouin zone path when plotting the periodic energy spectrum and the y-axis being the value of the real part of $\omega$ from 0.98$\omega_D$ to 1.02$\omega_D$, hawking all possible values of $\omega$ for the periodic energy spectrum.(b) The imaginary parts of the exact and Ewald series evolve as $\mathrm{N}(m,n=0,\pm1,... ,\pm \mathrm{N})$ varies, and since the Ewald series requires only a few terms to reach convergence, the $\mathrm{N}$ of it is fixed to 5. The black line denotes the Ewlad series, and the line marked by the blue small circles denotes the exact series. As $\mathrm{N}$ changes, the exact series keeps oscillating around the Ewald series and slowly converges to it.(c) The error (Err) between the imaginary part of the Ewald series and the exact series, the z-axis has log scale, and two points with large Err are marked randomly with red and green triangles respectively, the evolution of the series under the corresponding parameters are plotted in (b), (d), and the other settings are the same as in (a).(d) The same as (b) describes the evolution of the imaginary part of both series as the $\mathrm{N}$ varies.}
\end{minipage}
\end{figure}

Then discuss the dyadic part:
\begin{align}\label{smeq16}
\begin{split}
&\sum_{(m, n) \neq 0} \frac{1}{k^{2}} \nabla \otimes \nabla \frac{e^{-i k|\vec{R}|}}{4 \pi|\vec{R}|} \cdot e^{i \vec{q} \cdot \vec{R}} \\
&\approx \frac{1}{k^{2}}\left\{\left.\nabla \otimes \nabla G_{K}^{}\right|_{\left|\vec{r}_{u v}\right|=0}+\left.\nabla \otimes \nabla G_{R}^{}\right|_{(m, n) \neq 0,\left|\vec{r}_{u v}\right|=0}+\right. \\
&~~~~\left.\frac{1}{24 \pi}\left.\left(\frac{\partial^{3} f}{\partial R^{3}}\right|_{R=0}-2 i k^{3}\right) \cdot \hat{\mathbf{I}}\right\},
\end{split}
\end{align}
where $\nabla \otimes \nabla G_{R}$ and $\nabla \otimes \nabla G_{K}$ are the same as in Eq.(\ref{smeq4}), and the derivation of Eq.(\ref{smeq16}) also uses the regularization method \cite{A12}.Note also that the Eq.(\ref{smeq16}) is a $3 \times 3 $ matrix, in Z-modes here only take its (3,3) items, the specific method has been described in the previous text, here will not repeat. Eq.(\ref{smeq15}) plus Eq.(\ref{smeq16}) of the Z-modes part is the specific form of the Eq.(\ref{smeq14}).

To illustrate the validity of the previously derived Ewald series, it is compared here with the exact series of direct summation over a wide range of parameters.The parameters are set as follows: nanoparticle radius a=100nm, lattice constant d=15a, $\beta$=1.6 (in the range where the dipole approximation holds), $r_x=r_y=2c$. The Ewald series needs only a few terms to reach a great convergence, so here let $(m,n)=0,\pm1,\pm2,... ,\pm5$ (skip 0 if the summation sign is stated) \cite{B3}, a total of $11\times11=121$ terms need to be considered.For direct summation, here let $(m,n)=0,\pm1,\pm2,... ,\pm1000$ (skip 0 if the summation sign is stated), a total of $2001\times2001=4004001$ terms need to be considered. The final comparison results are shown in Fig.\ref{smfig1} and Fig.\ref{smfig2}.

Fig.\ref{smfig1} is the comparison result of $\Sigma \mathbf{G}[\begin{array}{l}\vec{R}+\vec{r}_{u v}\end{array}] \cdot t$ and Fig.\ref{smfig2} is the comparison result of $\Sigma^{\prime} \mathbf{G}(\vec{R}) \cdot t$.(a) and (c) of Fig.\ref{smfig1} and Fig.\ref{smfig2} indicate the comparison of the real and imaginary parts of the series, respectively, x-axis indicates the $q_x,q_y$ taking values along the high symmetry points, y-axis indicates the $\omega$ real part taking values from $0.98\omega_D$ to $1.02\omega_D$ (in the energy band of the dipole arrays, $\omega$ taking values generally in the $\ omega_D$ near) and the z-axis is the error (Err), which is specified by the expression \cite{A12}:
\begin{align}\label{smeq17}
\mathrm{Err}=\frac{|K^{\text{Ewald}}-K^{\text{Exact}}|}{|K^{\text{Exact}}|},
\end{align}
$K^{\text{Ewald}}$ denotes the Ewald series, and $K^{\text{Exact}}$ denotes the exact series of the direct summation.Observing (a) and (c) of Fig.\ref{smfig1} and \ref{smfig2}, it can be seen that the errors of the real part are within 5\text{\%}, and the z-axis of the imaginary part has a log scale. Most of the errors of the imaginary part are below 5\text{\%}, but some of them are large, which is because Eq.(\ref{smeq17}) does not reflect the actual errors truthfully in some cases. For example, when $K^{\text{Ewald}}=0$ and $K^{\text{Exact}}$ oscillates around 0 leads to the wrong conclusion that the $\mathrm{Err}=1$, or when $K^{\text{Ewald}}$ converges around 0 and is larger than 0, while $K^{\text{Exact}}$ is around 0 but less than 0 leads to the wrong conclusion that the $\mathrm{Err}>1$.To observe the actual convergence effect, any two points with large $\mathrm{Err}$ (more than 5\%) are selected in (c) of Fig.\ref{smfig1} and Fig.\ref{smfig2} respectively and marked with triangles of different colors. For the selected parameters, the changes of Ewald series and exact series with $\mathrm{N}(m,n=0,\pm1,... ,\pm \mathrm{N})$ are plotted in (b), (d) of Fig.\ref{smfig1} and \ref{smfig2} respectively (note that the $\mathrm{N}$ of the Ewald series is fixed to 5).It can be seen that as $\mathrm{N}$ becomes larger, the exact series keeps oscillating up and down with the Ewald series as the center and slowly converges to it \cite{A52}, which shows that the Ewlad series constructed in this section has well convergence and can accurately describe the full coupling of the system.

\subsection{Mixed boundary conditions}
The lattice sums technique constructed under mixed boundary conditions is somewhat more complicated than that under periodic boundary conditions, which not only utilizes the 1D Ewlad method \cite{A09}, but also incorporates the transcendent function method \cite{A40} in order to simplify the discussion. In practice, it is found that it also has certain requirements on the aspect ratio of nanoribbons, which are discussed in detail in this section. As in the case of the periodic boundary conditions, the elements in the eigenmatrix of the nanoribbon are mostly repeated, and only two items need to be considered here:$\sum_{n \in Z} \mathbf{G}(n d \cdot \hat{y}+\vec{r}) \cdot e^{i q_y \cdot n d}$, $\sum_{n \neq 0} \mathbf{G}(n d \cdot \hat{y})\cdot e^{i q_y \cdot n d}$, $\vec{r}=r_x \cdot \hat{x}+r_y \cdot \hat{y}+r_z \cdot \hat{z}$, which has a meaning similar to that of $\vec{r}_{uv}$ in the periodic boundary conditions, it denotes a vector of non-zero size between any two lattice points in the same nanoribbon ($r_z$ is set to 0 at the end).

Discussing first $\sum_{n \in Z} \mathbf{G}(n d \cdot \hat{y}+\vec{r}) \cdot e^{i q_y \cdot n d}$, as before, the following formula can be divided into a scalar part and a dyadic part:
\begin{align}\label{smeq18}
\begin{split}
&\sum_{n \in Z} \mathbf{G}(n d \cdot \hat{y}+\vec{r}) \cdot e^{i q_y \cdot n d} \\
&=\sum_{ n \in Z}\left\{\hat{\mathbf I} \cdot \frac{e^{-i k\left|n d \cdot \hat{y}+\vec{r}\right|}}{4 \pi\left|n d \cdot \hat{y}+\vec{r}\right|} \cdot e^{i q_y \cdot n d}+\frac{1}{k^{2}} \nabla \otimes \nabla\left(\frac{e^{-i k\left|n d \cdot \hat{y}+\vec{r}\right|}}{4 \pi\left|n d \cdot \hat{y}+\vec{r}\right|} \cdot e^{i q_y \cdot n d}\right)\right\},
\end{split}
\end{align}
Eq.(\ref{smeq18}) needs to be discussed in the cases of $r_x\neq 0$ and $r_x= 0$.

When $r_x\neq0$,the scalar part using the 1D Ewald method can be written as\cite{A04,A08,A09}:
\begin{align}\label{smeq19}
\begin{split}
&\sum_{n\in Z}\frac{e^{-i k\left|n d \cdot \hat{y}+\vec{r}\right|}}{4 \pi\left|n d \cdot \hat{y}+\vec{r}\right|} \cdot e^{i q_y \cdot n d}\\
&=\sum_{n \in Z} \frac{e^{-i k \cdot R r_{ n}}}{4 \pi R r_{n}} \cdot e^{i q_y \cdot n d}\\
&=G_{R,\mathbf{1D}}+G_{K,\mathbf{1D}}\\
&=\frac{1}{8 \pi} \sum_{n \in Z} \frac{e^{i q_y \cdot n d}}{R r_{n}}\left\{\sum_{\pm} e^{\pm i k \cdot R r_{n}} \operatorname{erfc}\left(R r_{ n} \cdot E \pm \frac{i k}{2 E}\right)\right\}_{\text{R}}+\\
&~~~~~\frac{1}{4 \pi d} \sum_{n\in Z} e^{i K_Y \cdot r_y}  \left\{\sum_{p=0}^{\infty} \frac{(-1)^p}{p !}\left(\sqrt{r_x^2+r_z^2} \cdot E\right)^{2 p} \cdot \mathbf{E}_{p+1}\left(\frac{\Gamma_n^2}{4 E^2}\right)\right\}_{\text{K}},
\end{split}
\end{align}
where $Rr_n=\left|n d \cdot \hat{y}+\vec{r}\right|$, which is a scalar of non-zero size that represents the distance between any two lattice points in the structure under mixed boundary conditions, and $q_y$ is the magnitude of the Bloch wave vector in the y-direction, where $q_y \in[-\pi/d,\pi/d]$.$\Gamma_n^2=K_Y^2-k^2$,$k$ has the same meaning as in the periodic condition, $K_Y={2 \pi n}/{d}-q_y$. The $\mathbf{E}_p(\cdot)$ is the $p$ order exponential integral \cite{A54}.
$E$ is the splitting parameter, $ E=\max(\sqrt{{\pi}/{d}},~{\mathrm{Re}(k)}/{2 H})$, and the empirical parameter $H$ is still set to 3 \cite{A08}.

Then discuss the dyadic part:
\begin{align}\label{smeq20}
\begin{split}
&\sum_{n\in Z}\frac{1}{k^{2}} \nabla \otimes \nabla\left(\frac{e^{-i k\left|n d \cdot \hat{y}+\vec{r}\right|}}{4 \pi\left|n d \cdot \hat{y}+\vec{r}\right|} \cdot e^{i q_y \cdot n d}\right)\\
&=\frac{1}{k^{2}}\left\{\nabla \otimes \nabla G_{K,\mathbf{1D}}+\nabla \otimes \nabla G_{R,\mathbf{1D}}\right\},
\end{split}
\end{align}
as before, Eq.(\ref{smeq20}) is divided into the dyadic of real space series and the k-space series. Referring to the previous steps, the dyadic of real space series in Z-modes can be written directly as:
\begin{align}\label{smeq21}
\begin{split}
&\left.\frac{1}{k^2} \nabla \otimes \nabla G_{R,\mathbf{1D}}\right |_z \\
&=\frac{1}{k^2}\cdot \frac{1}{8 \pi} \sum_{n \in Z} e^{i q_y \cdot nd} \cdot\left(\frac{\partial f}{\partial R r_{ n}} \cdot \frac{1}{R r_{ n}^2}-\frac{f}{R r_{ n}^3}\right),
\end{split}
\end{align}
the expressions for each order derivative of $f$ and $f$ are the same as in the periodic boundary conditions, with the difference that $Rr_{mn}$ needs to be replaced by $Rr_n$.

The dyadic of the k-space series is a matrix of size $3\times 3$, and only its (3,3) term is considered in the Z-modes:
\begin{align}\label{smeq22}
\begin{split}
&\left.\frac{1}{k^2} \nabla \otimes \nabla G_{K,\mathbf{1D}}\right|_z\\
&=\partial_z\nabla G_{K,\mathbf{1D}}^z \\
&=G_{K,\mathbf{1D}} \cdot\left(\frac{2 p \cdot r_z}{r_x^2+r_z^2}\right)^2+G_{K,\mathbf{1D}} \cdot\left(\frac{2 p}{r_x^2+r_z^2}-4 p \cdot \frac{r_z^2}{\left(r_x^2+r_z^2\right)^2}\right),
\end{split}
\end{align}
where $\nabla G_{K,\mathbf{1D}}^z$ is the z-component of the $\nabla G_{K,\mathbf{1D}}$ vector. It should also be noted that the meaning of $p$ in Eq.(\ref{smeq22}) is the same as that of $p$ in Eq.(\ref{smeq19}), which are both summation parameters.

When $r_x=0$:
\begin{align}\label{smeq23}
\begin{split}
&\left.\sum_{n \in Z} \mathbf{G}\left(n d \cdot \hat{y}+\vec{r}\right) \cdot e^{i q_y \cdot n d}\right|_z \\
&=\sum_{n \in Z} \frac{e^{-i k\left|n d+r_y\right|}}{4 \pi\left|n d+r_y\right|}\left\{1-\frac{i k\left|n d+r_y\right|+1}{k^2\left|n d+r_y\right|^2}\right\} \cdot e^{i q_y \cdot n d},
\end{split}
\end{align}
Eq.(\ref{smeq23}) needs to be divided into $r_y>0$ and $r_y<0$ to discuss separately, using the transcendent function method, when $r_y>0$, it is obtained that:
\begin{align}\label{smeq24}
\begin{split}
&\left.\sum_{n \in Z} \mathbf{G}\left(n d \cdot \hat{y}+\vec{r}\right) \cdot e^{i q_y \cdot n d}\right|_z\\
&= \frac{e^{-i k \cdot r_y}}{4 \pi d} \cdot \Phi\left[e^{i d \cdot\left(q_y-k\right)}, 1, \frac{r_y}{d}\right]-\\
&~~~~\frac{i \cdot e^{-i k \cdot r_y}}{4 \pi k \cdot d^2} \cdot \Phi\left[e^{i d\left(q_y-k\right)}, 2, \frac{r_y}{d}\right]-\\
& ~~~~\frac{e^{-i k \cdot r_y}}{4 \pi k^2 d^3} \cdot \Phi\left[e^{i d \cdot\left(q_y-k\right)} ,3, \frac{r_y}{d}\right]+\\
&~~~~ \frac{e^{-i d \cdot\left(k+q_y\right)} \cdot e^{i k \cdot r_y}}{4 \pi d} \cdot \Phi\left[e^{-i d\left(k+q_y\right)}, 1,1-\frac{r_y}{d}\right]-\\
&~~~~\frac{i \cdot e^{-i d \cdot\left(k+q_y\right)} \cdot e^{i k \cdot r_y}}{4 \pi k d^2} \cdot \Phi\left[e^{-i d\left(k+q_y\right)}, 2,1-\frac{r_y}{d}\right]-\\
&~~~~\frac{e^{-i d\left(k+q_y\right)} \cdot e^{i k \cdot r_y}}{4 \pi k^2 d^3} \cdot \Phi\left[e^{-i d\left(k+q_y\right)}, 3,1-\frac{r_y}{d}\right],
\end{split}
\end{align}
when $r_y<0$, it is obtained that:
\begin{align}\label{smeq25}
\begin{split}
&\left.\sum_{n \in Z} \mathbf{G}\left(n d \cdot \hat{y}+\vec{r}\right) \cdot e^{i q_y \cdot n d}\right|_z \\
&= \frac{e^{i k \cdot r_y}}{4 \pi d} \cdot \Phi\left[e^{-i d\left(k+q_y\right)}, 1,-\frac{r_y}{d}\right]-\\
&~~~~ \frac{i \cdot e^{i k \cdot r_y}}{4 \pi k \cdot d^2} \cdot \Phi\left[e^{-i d\left(k+q_y\right)}, 2,-\frac{r_y}{d}\right]-\\
&~~~~\frac{e^{i k \cdot r_y}}{4 \pi k^2 d^3} \cdot \Phi\left[e^{-i d\left(k+q_y\right)}, 3,-\frac{r_y}{d}\right]+\\
&~~~~ \frac{e^{i d\left(q_y-k\right)} \cdot e^{-i k \cdot r_y}}{4 \pi d} \cdot \Phi\left[e^{i d\left(q_y-k\right)}, 1,1+\frac{r_y}{d}\right]-\\
&~~~~ \frac{i \cdot e^{i d\left(q_y-k\right)} \cdot e^{-i k \cdot r_y}}{4 \pi k d^2} \cdot \Phi\left[e^{i d\left(q_y-k\right)}, 2,1+\frac{r_y}{d}\right]-\\
&~~~~ \frac{e^{i d\left(q_y-k\right)} \cdot e^{-i k \cdot r_y}}{4 \pi k^2 d^3} \cdot \Phi\left[e^{i d\left(q_y-k\right)}, 3,1+\frac{r_y}{d}\right],
\end{split}
\end{align}
where $\Phi(\cdot)$ is called the Lerch transcendent, and the detailed definition can be found in \cite{A40,A54}.

In conclusion when $r_x \neq 0$, the accelerated convergence of Eq.(\ref{smeq18}) takes the form of Eqs.(\ref{smeq19}), (\ref{smeq21}), (\ref{smeq22}); when $r_x = 0$, the accelerated convergence of Eq.(\ref{smeq18}) takes the form of Eq.(\ref{smeq24}) or Eq.(\ref{smeq25}).

Then discuss $\sum_{n \neq 0} \mathbf{G}(n d \cdot \hat{y})\cdot e^{i q_y \cdot n d}$, which reduces in the Z-modes to:
\begin{align}\label{smeq26}
\begin{split}
&\left.\sum_{n \neq 0} \mathbf{G}(n d \cdot \hat{y}) \cdot e^{i q_y \cdot n d}\right|_z \\
&=\sum_{n \neq 0} \frac{e^{-i k|n d|}}{4 \pi|n d|}\left\{1-\frac{i k|n d|+1}{k^2|n d|^2}\right\} \cdot e^{i q_y \cdot n d},
\end{split}
\end{align}
\begin{figure}[!htbp]
\begin{minipage}{0.49\textwidth}
\centering
\includegraphics[width=8cm]{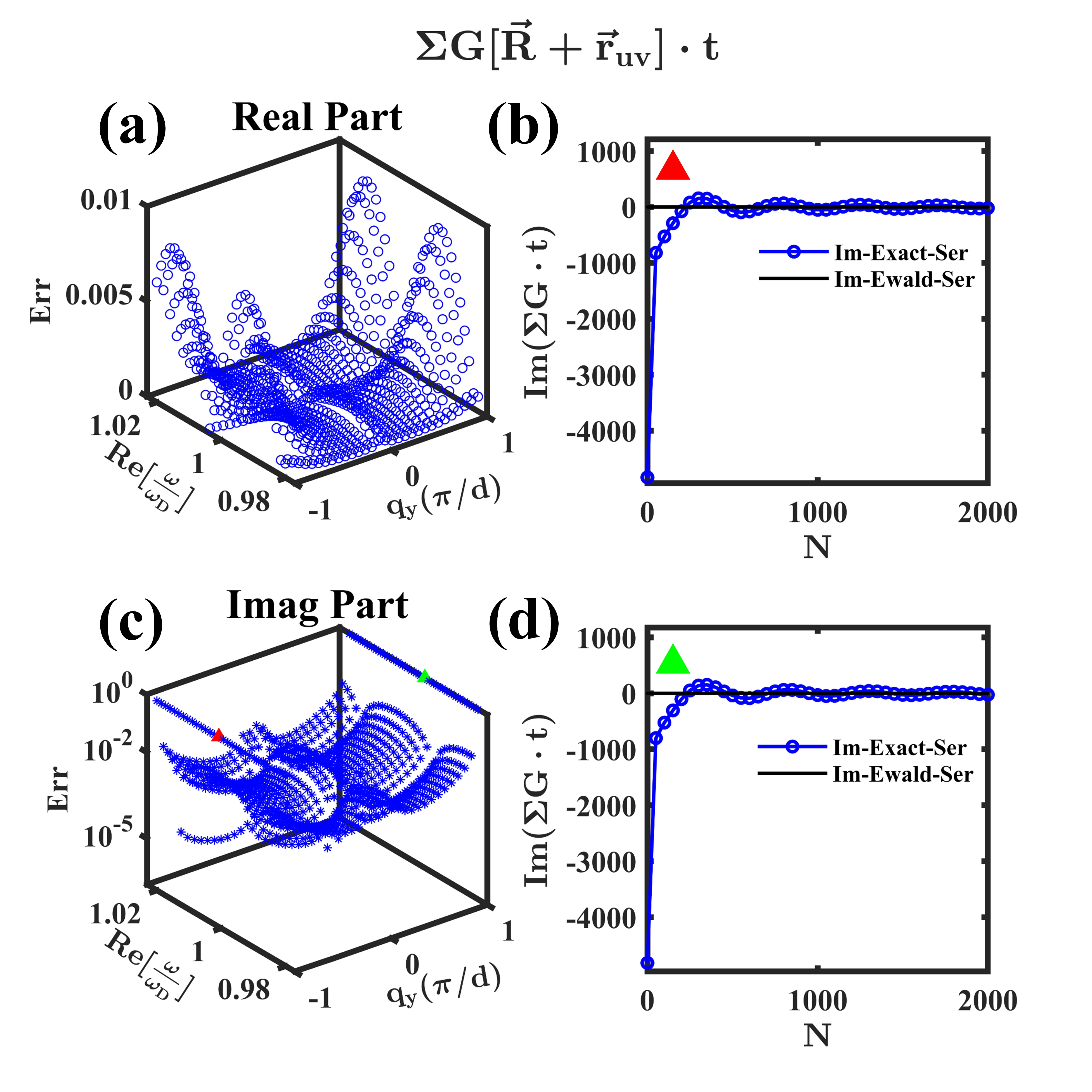}
\caption{\label{smfig3}Comparison of the direct summation of $\sum_{n \in Z} \mathbf{G}(n d \cdot \hat{y}+\vec{r}) \cdot e^{i q_y \cdot n d}$ (exact series) with the lattice sums technique.(a) Error (Err) between the real part of the lattice sums technique and the exact series, with the x-axis represents the value of $q_y$ and the y-axis being the value of the real part of $\omega$ from 0.98$\omega_D$ to 1.02$\omega_D$, hawking all possible values of $\omega$ for the  energy spectrum.(b) The imaginary parts of the exact series and lattice sums technique evolve as $\mathrm{N}(m,n=0,\pm1,... ,\pm \mathrm{N})$ varies, and since the lattice sums technique requires only a few terms to reach convergence, the $\mathrm{N}$ of it is fixed to 5. The black line denotes the lattice sums technique, and the line marked by the blue small circles denotes the exact series. As $\mathrm{N}$ changes, the exact series keeps oscillating around the lattice sums technique and slowly converges to it.(c) The error (Err) between the imaginary part of the lattice sums technique and the exact series, the z-axis has log scale, and two points with large Err are marked randomly with red and green triangles respectively, the evolution of the series under the corresponding parameters are plotted in (b), (d), and the other settings are the same as in (a).(d) The same as (b) describes the evolution of the imaginary part of both series as the $\mathrm{N}$ varies.}
\end{minipage}
\hfill
\begin{minipage}{0.49\textwidth}
\centering
\includegraphics[width=8cm]{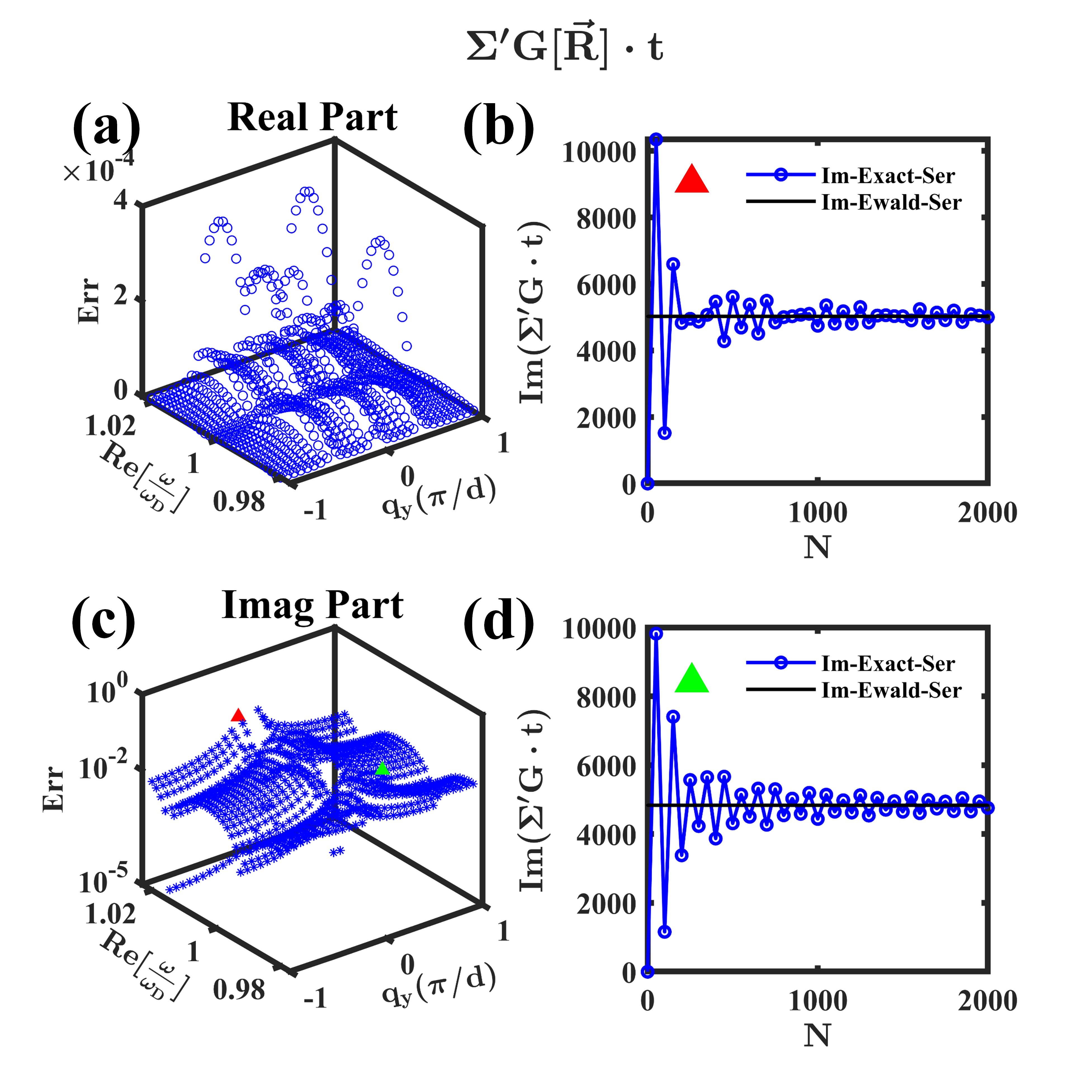}
\caption{\label{smfig4}Comparison of the direct summation of $\sum_{n \neq 0} \mathbf{G}(n d \cdot \hat{y})\cdot e^{i q_y \cdot n d}$ (exact series) with the lattice sums technique.(a) Error (Err) between the real part of the lattice sums technique and the exact series, with the x-axis represents the value of $q_y$ and the y-axis being the value of the real part of $\omega$ from 0.98$\omega_D$ to 1.02$\omega_D$, hawking all possible values of $\omega$ for the  energy spectrum.(b) The imaginary parts of the exact series and lattice sums technique evolve as $\mathrm{N}(m,n=0,\pm1,... ,\pm \mathrm{N})$ varies, and since the lattice sums technique requires only a few terms to reach convergence, the $\mathrm{N}$ of it is fixed to 5. The black line denotes the lattice sums technique, and the line marked by the blue small circles denotes the exact series. As $\mathrm{N}$ changes, the exact series keeps oscillating around the lattice sums technique and slowly converges to it.(c) The error (Err) between the imaginary part of the lattice sums technique and the exact series, the z-axis has log scale, and two points with large Err are marked randomly with red and green triangles respectively, the evolution of the series under the corresponding parameters are plotted in (b), (d), and the other settings are the same as in (a).(d) The same as (b) describes the evolution of the imaginary part of both series as the $\mathrm{N}$ varies.}
\end{minipage}
\end{figure}
Using the transcendent function method, Eq.(\ref{smeq26}) is written as:
\begin{align}
\begin{split}
&\left.\sum_{n \neq 0} \mathbf{G}(n d \cdot \hat{y}) \cdot e^{i q_y \cdot n d}\right|_z\\
&=\frac{1}{4 \pi d}\left\{\mathrm{Li}_1\left(e^{i\left(q_y-k\right) d}\right)+\mathrm{Li}_1\left(e^{-i\left(k+q_y\right) d}\right)\right\} \\
&~~~~~-\frac{i}{4 \pi k \cdot d^2}\left\{\mathrm{Li}_2\left(e^{i\left(q_y-k\right) d}\right)+\mathrm{Li}_2\left(e^{-i\left(k+q_y\right) d}\right)\right\} \\
&~~~~~-\frac{1}{4 \pi k^2 d^3}\left\{\mathrm{Li}_3\left(e^{i(q_y-k) d}\right)+\mathrm{Li}_3\left(e^{-i\left(k+q_y\right) d}\right)\right\},
\end{split}
\end{align}
where $\mathrm{Li}_{s}(\cdot)$ is called the polylogarithms function, and a detailed definition can be found in \cite{A54}.

Overall, since the transcendent function method does not accelerate $\sum_{n \in Z} \mathbf{G}(n d \cdot \hat{y}+\vec{r}) \cdot e^{i q_y \cdot n d}$ for the $r_x \neq0$ case, it is necessary to use the 1D Ewald method in the mixed boundary conditions,but the transcendent function method can also simplify the discussion of the problem, for example, by avoiding the use of regularization methods, so the two approaches are complementary under mixed boundary conditions.

It is found in practice that the 1D Ewald series form of $\sum_{n \in Z} \mathbf{G}(n d \cdot \hat{y}+\vec{r}) \cdot e^{i q_y \cdot n d}$ has a range of use, where the lattice constant of the nanoribbon supercell is less than or equal to $60d$ (here let $d=15a$), the maximum value of $r_x$ should be no more than three times the lattice constant of the supercell, that is, there is a requirement for the ratio between the lengths of the nanoribbon in the x-direction and the y-direction.
Also in practice, it is found that setting the summation parameter $p$ to about 100 will have a better convergence effect, less than 100 is too much then the Ewald series does not converge, more than 100 is too much then it will lead to $G_{K,\mathbf{1D}}$ and $\nabla \otimes \nabla G_{K,\mathbf{1D}}$ with serious precision loss issues thus leading to NaN.
Since it is beyond the scope of this research, these issues are not analyzed in detail in this paper.

To illustrate the validity of the previously derived lattice sums technique, here it is compared with the exact series of direct summation over a large range of parameters.The parameters are set as follows: nanoparticle radius $a=100 \text{nm}$, lattice constant $d=15a$, $\beta=0.4$, $r_x=2d+2c$, $r_y=2c$. Also to ensure the convergence effect, for the lattice sums technique, here $n=0,\pm1,... ,\pm5$, $p=100$, the lattice constant of the nanoribbon supercell is set to $d$ (i.e., the height in the period direction is $d$), and the length in the x-direction (open boundary direction) is set to $3d$.For direct summation, here let $n=0,\pm1,... ,\pm1000$ (skipping 0 if the summation sign is specified). The final comparison results are shown in Fig.\ref{smfig3} and \ref{smfig4}.

Fig.\ref{smfig3} is the comparison result of $\sum_{n \in Z} \mathbf{G}(n d \cdot \hat{y}+\vec{r}) \cdot e^{i q_y \cdot n d}$ and Fig.\ref{smfig4} is the comparison result of $\sum_{n \neq 0} \mathbf{G}(n d \cdot \hat{y})\cdot e^{i q_y \cdot n d}$.
(a) and (c) of Fig.\ref{smfig3} and \ref{smfig4} represent the comparison of the real and imaginary parts of the series. The x-axis represents the value of $q_y$, the y-axis represents the value of the real part of $\omega$ from $0.98\omega_D$ to $1.02\omega_D$ (in the energy band of the dipole arrays, the value of $\omega$ is usually taken around $\omega_D$ ), and the z-axis is the error (Err), which has the same specific expression as before. The errors presented in (a) for both the Fig.\ref{smfig3} and \ref{smfig4} maps are less than 5\%.And the Err here as before does not faithfully reflect the error between the two series in some cases. So similarly, to observe the actual convergence effect, any two points with large $\mathrm{Err}$ (more than 5\%) are selected and marked with triangles of different colors in the (c) plots of Fig.\ref{smfig3} and \ref{smfig4} respectively, and then the variation of the lattice sums technique and the exact series with the $\mathrm{N}(m, n=0,\pm1,..., \pm \mathrm{N})$ are drawn in (b), (d) of Fig.\ref{smfig3} and \ref{smfig4} respectively (it should be noted that the $\mathrm{N}$ of the lattice sums technique are fixed to 5). It can be seen that as $\mathrm{N}$ becomes larger, the exact series keeps oscillating and slowly converges to the results derived from the lattice sums technique, which indicates that the lattice sums technique constructed in this section has excellent convergence in the established range and can accurately describe the full coupling of the system.

The main text utilizes the truncated summation ($n=0,\pm1,. ,\pm1000$) to draw the projected bands. To illustrate its accuracy, the band diagrams of the nanoribbon with length $d$ in the y-direction and $3d(d=15a)$ in the x-direction are drawn using the lattice sums technique in Appendix \ref{smsec3}, provided that the lattice sums technique holds. It can be seen from Fig.\ref{smfig6} that it presents completely the same topological properties as in the main text. Also if want to draw the energy spectrum of a nanoribbon with the length in the x-direction (open boundary) exceeding $3d$, the ratio of its length in the y-direction (periodic boundary) to the length in the x-direction should be controlled to be within the range described earlier.

\section{\label{smsec2}DEGENERATE BANDS’ WANNIER BAND }
This section plots the wannier bands for the second and third bulk band (degenerate bands), using Eq.(\ref{eq10}) in the main text.
As shown in Fig.\ref{smfig5}, $d=15a$ and $\beta$ is 0.4, 0.99, 1.01, 1.6 from left to right, it can be seen that the wannier bands from (a) to (d) is either near 0 or near $\pm0.5$, and there is a phase transition from 0.99 to 1.01, these are the same topological properties presented by the bulk polarization in Fig.4(a) of the main text .
\begin{figure}[!htbp]
\centering
\includegraphics[width=0.85\linewidth]{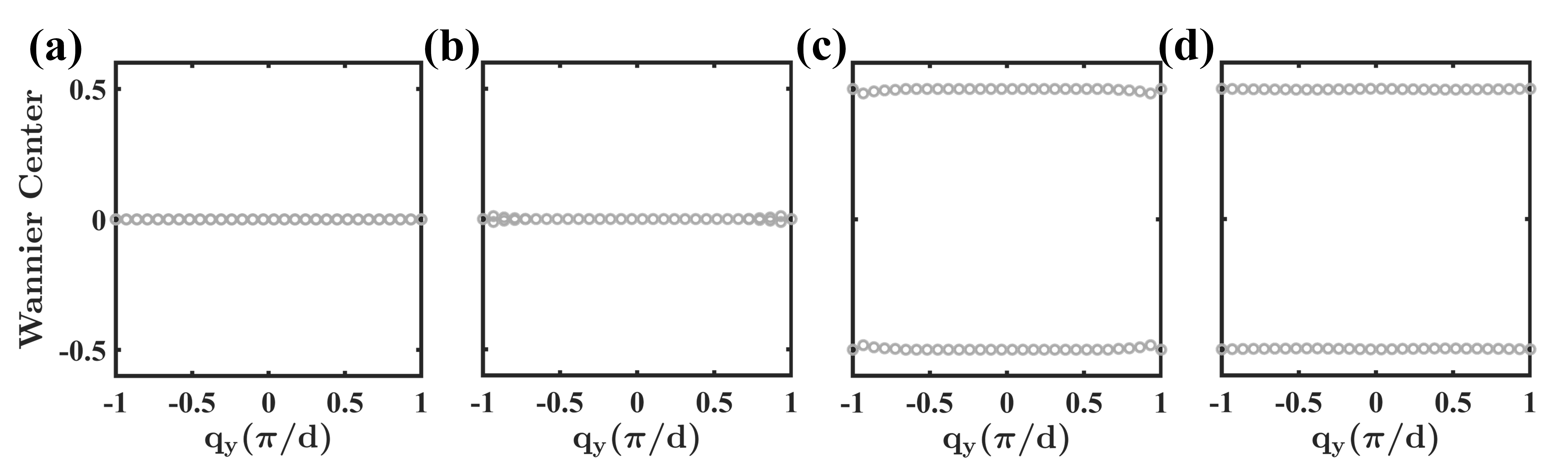}
\caption{\label{smfig5}The wannier bands of the second and third bands of the bulk spectrum, $d=15a$, $q_y\in[-\pi/d,\pi/d]$. (a)-(d)$\beta$ are set to 0.4, 0.99, 1.01, 1.6. Their wannier centers are localized to 0 or $\pm0.5$.}
\end{figure}

\section{\label{smsec3}MAPPING THE BAND STRUCTURE OF NANORIBBONS: BASED UPON THE LATTICE SUMS TECHNIQUE}
This section plots the projection bands of the nanoribbons using the lattice sums technique for the mixed boundary conditions derived in the Appendix \ref{smsec1}. As shown in Fig.\ref{smfig6}, $d=15a$, the length of the nanoribbon in the x-direction (open boundary) is 3$d$, the length in the y-direction (periodic boundary) is $d$, the $\beta$ of (a) to (d) are set to 0.7, 1.1, 1.2, 1.3, and the $\beta$ is set in the identical range as in Fig.2 of the main text.The bands marked in red in the figure indicate the edge states, and the edge states above and below the zero-energy are two-fold degenerate. $\beta=1.1$ when the two ends of the edge state and the bulk are mixed together, and when $\beta=1.2$ a completely discrete edge state appears, which indicates that the stable edge state appears between 1.1 and 1.2. Also the gray dashed line indicates zero-energy, $\beta=1.2$ when the dashed line and the bulk intersect, and $\beta=1.3$ when them do not intersect, which means that the robust off-bulk corner states appears between 1.2 and 1.3. The above conclusion is consistent with that obtained from the projected bands plotted using truncated summation in the main text, which illustrates the validity of truncated summation.
\begin{figure}[!htbp]
\centering
\includegraphics[width=0.85\linewidth]{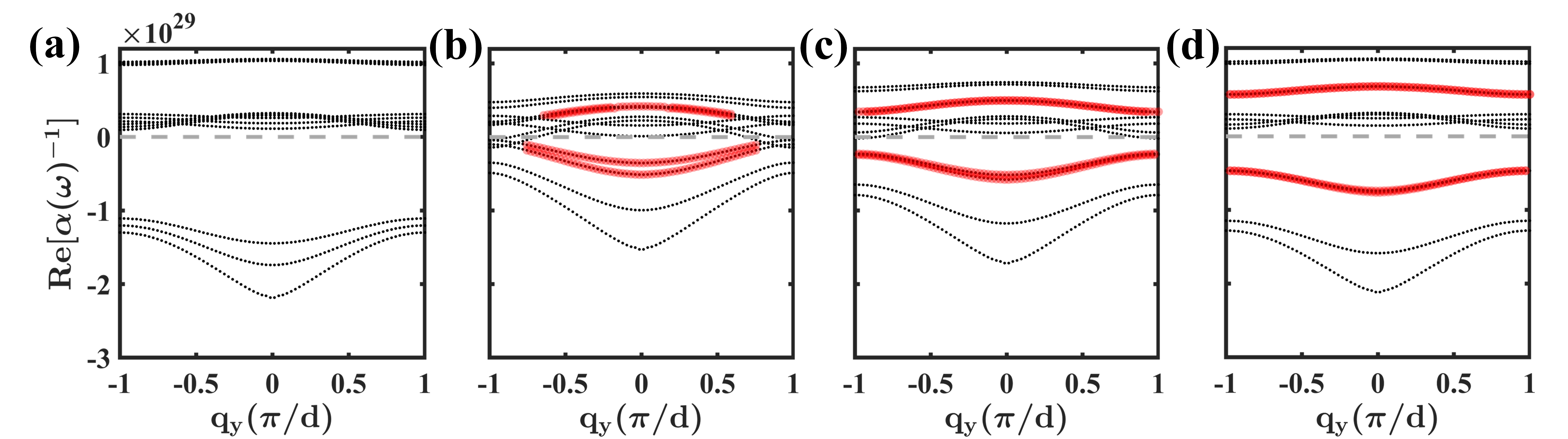}
\caption{\label{smfig6}Projected bands of the nanoribbon drawn using the lattice sums technique, with the nanoribbon having lengths of 3$d$ in the x-direction (open boundary) and $d$ in the y-direction (periodic boundary), $d=15a$, $q_y \in[-\pi/d,\pi/d]$.(a)-(d)$\beta$ are set to 0.7, 1.1, 1.2, and 1.3, respectively, and the bands marked in red indicate the edge states, and the edge states above and below the zero-energy are two-fold degenerate, and the gray dashed line indicates the case where the y-axis is 0.}
\end{figure}

\section{\label{smsec4}TOPOLOGICAL PROPERTIES AND THE EIGEN-RESPONSE THEORY}
In this section, the topological properties of the open boundary structure (20 UCs in both x- and y-directions) in Sec.IV of the main text are re-examined using the so-called eigen-response theory \cite{A37,A43,A53}.Unlike Eq.(\ref{eq5}) in the main text, which fixes the RHS $\omega$ as $\omega_D$ (linearization), here all $\omega$ are considered as variables and made to vary around $\omega_D$. Specifically, with the idea of Eq.(\ref{eq5}) in the main text, an eigenmatrix $\mathcal{M}(\omega)_{1600 \times 1600}$ of size $1600 \times 1600$ is constructed, followed by:
\begin{align}
\begin{split}
\mathcal{G}(\omega)=\frac{1}{\alpha(\omega)}\hat{\mathbf{I}}_{1600\times 1600}-\frac{k^2}{\epsilon _0}\mathcal{M}(\omega)_{1600\times 1600},
\end{split}
\end{align}
where $\alpha(\omega)$ has the same specific form as Eq.(\ref{eq3}) in the main text, $\hat{\mathbf{I}}$ is the unit matrix, and the DGF for constructing the matrix $\mathcal{M}_{1600\times1600}$ is:$\textstyle \mathbf{G}=\frac{e^{-i k R}}{4 \pi R}\left\{1-\frac{i k R+1}{k^2 R^2}\right\}$.

Let $\lambda$ be the complex eigenvalues of $\mathcal{G}(\omega)$, while having:
\begin{align}
\alpha_{eig}=\frac{1}{\lambda},
\end{align}
the $\alpha_{eig}$ is called eigen-polarizability, and different $\omega$ correspond to different $\alpha_{eig}$.Here let $\omega$ vary from 0.997$\omega_D$ to 1.003$\omega_D$, and draw $\mathrm{Im}(\alpha_{eig})$ under different $\omega$ correspondences and thus obtained Fig.\ref{smfig7}.From (a)-(f), $\beta$ is set to 0.8, 0.9, 1, 1.1, 1.2, 1.3 respectively, and d = 15a.
\begin{figure}[!htbp]
\centering
\includegraphics[width=0.85\linewidth]{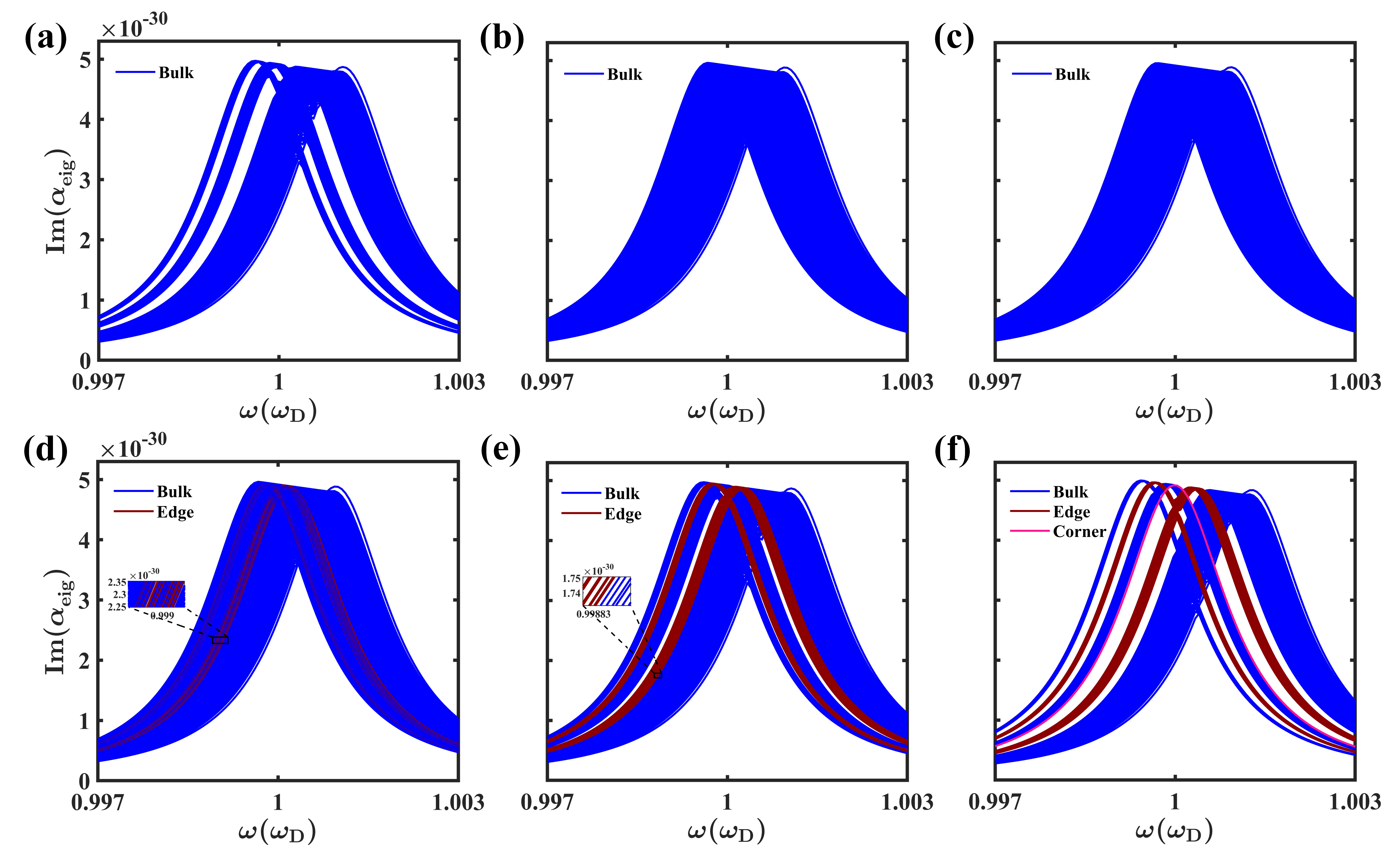}
\caption{\label{smfig7} The eigen-polarizability of the open boundary structure with 20 UCs in both x- and y-directions in the Z-modes. The $\omega$ changes from 0.997$\omega_D$ to 1.003$\omega_D$, $d=15a$. Blue indicates bulk states,dark red indicates edge states, deep pink indicates corner states. (a)-(f)$\beta$ are set to 0.8, 0.9, 1, 1.1, 1.2, 1.3 respectively.}
\end{figure}

The blue color in Fig.\ref{smfig7} indicates the bulk states, the dark red indicates the edge states, and the deep pink indicates the corner states.The system has no topological states when $\beta=$0.8, 0.9, 1. When $\beta$ = 1.1 the edge and bulk states are mixed together as seen in the enlarged view in (d).Also there are BIC under this parameter, but they are not marked here due to their instability.At $\beta$ = 1.2 the edge states and the bulk states are separated as seen by the enlarged view in (e), which means that the stable off-bulk edge states appear between 1.1 and 1.2. The corner states appear at $\beta$ = 1.3, which implies that the stable robust corner states appear between 1.2 and 1.3. The topological properties presented above are exactly the same as those in the fixed $\omega$ case, indicating that the research under the linearization assumption can accurately capture the topological properties of the variable $\omega$ system.

\section{\label{smsec5}THE STANDARD BAND STRUCTURE }

The main text treats $1/(\alpha(\omega))$ as eigenvalues for simplicity the discussion, which is the same as the topological properties presented by the energy spectrum with $\omega$ as eigenvalues. To illustrate this point, the energy spectrum with $\omega$ as the eigenvalue is plotted in this section in conjunction with Eq.(\ref{eq3}) in the main text, where the parameters of the periodic structure are $d=15a$ and $\beta=1.3$, and the parameters of the open boundary structure with 20 UCs in both x- and y-directions are the same as those of the periodic structure.In a strict sense, the system studied in this paper is non-Hermitian, so the imaginary part of the bands with the corresponding parameters is also drawn here.

As shown in Fig.\ref{smfig8}, (a) (b) are the real and imaginary parts of the periodic energy spectrum respectively, and (c) (d) are the real and imaginary parts of the open boundary energy spectrum respectively, the imaginary part is generally three orders of magnitude smaller than the real part. The black dashed line is the position of $\omega=\omega_D$ and the gray dashed line indicates the light cone \cite{A33}. Except for the differences in geometric positions, Fig.\ref{smfig8} presents the exact same topological properties as described in the main text.
\begin{figure}
\centering
\includegraphics[width=0.7\linewidth]{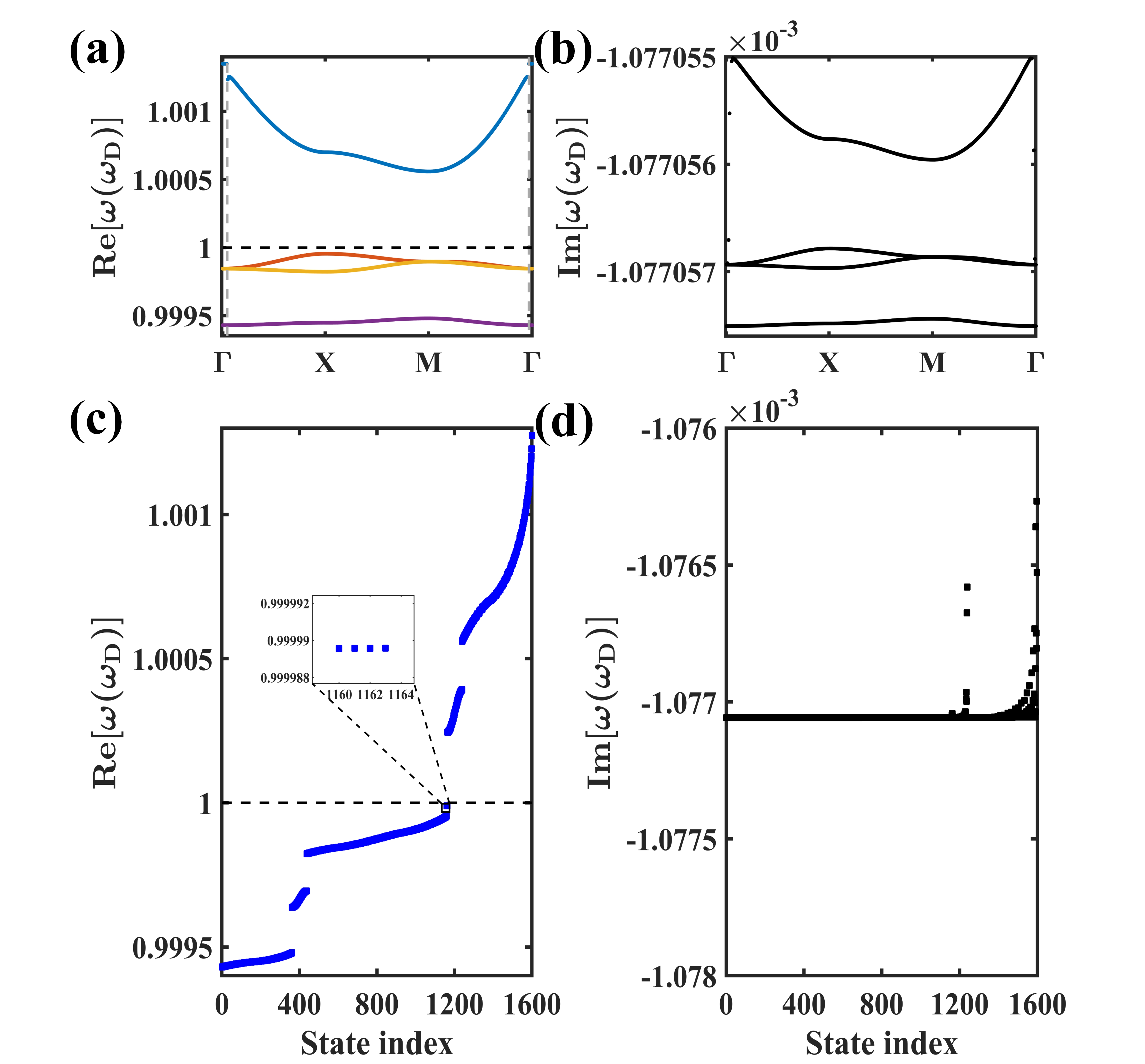}
\caption{\label{smfig8}The standard energy spectrum with $\omega$ as the eigenvalues. (a) (b) The real and imaginary parts of the periodic energy spectrum with parameters $d=15a$ and $\beta=1.3$ respectively. (c) (d) The real and imaginary parts of the open boundary energy spectrum respectively, with the structure size of $20\times 20$ UCs, and other parameters are the same as those of the periodic energy spectrum. The black dashed line is the position of $\omega=\omega_D$, and the gray dashed line indicates the light cone. }
\end{figure}
\end{widetext}

\newpage
\bibliography{Ref_Me}

\end{document}